\documentclass[letterpaper,showpacs,onecolumn,eqsecnum,superscriptaddress,nofootinbib]{revtex4}

\usepackage{amssymb}
\usepackage{amsmath}
\usepackage{latexsym}
\usepackage{mathtools}
\DeclarePairedDelimiter\abs{\lvert}{\rvert} 
\usepackage{tensor} 
\usepackage{framed}
\makeatletter
\newcommand{\mathleft}{\@fleqntrue\@mathmargin0pt}
\newcommand{\mathcenter}{\@fleqnfalse}
\makeatother
\usepackage{epsfig}
\usepackage{epstopdf}
\usepackage{graphicx}
\usepackage{subcaption}
\AtBeginDocument{
\cmidrulewidth=.03em
\cmidrulesep=\doublerulesep
\cmidrulekern=.5em
}

\usepackage{color}

\newtheorem{theorem}{Theorem}
\newcommand{\proof}{\noindent {\bf Proof. }}
\newcommand{\qed}{\hfill $\fbox{\hspace{0.3mm}}$ \vspace{.3cm}} 

\usepackage{booktabs}

\begin{document}

\title{Traversable $\ell$-wormholes supported by ghost scalar fields}

\author{Belen Carvente} 
\email[]{belen.carvente@correo.nucleares.unam.mx} 
\affiliation{Instituto de Ciencias Nucleares, Universidad Nacional
  Aut\'onoma de M\'exico, Circuito Exterior C.U., A.P. 70-543,
  M\'exico D.F. 04510, M\'exico}
  
\author{V\'ictor Jaramillo} 
\email[]{victor.jaramillo@correo.nucleares.unam.mx} 
\affiliation{Instituto de Ciencias Nucleares, Universidad Nacional
  Aut\'onoma de M\'exico, Circuito Exterior C.U., A.P. 70-543,
  M\'exico D.F. 04510, M\'exico}

\author{Juan Carlos Degollado} 
\email[]{jcdegollado@icf.unam.mx}
\affiliation{Instituto de Ciencias F\'isicas, Universidad Nacional Aut\'onoma 
de M\'exico, Apdo. Postal 48-3, 62251, Cuernavaca, Morelos, M\'exico}

\author{Dar\'{\i}o N\'u\~nez}
\email[]{nunez@nucleares.unam.mx}
\affiliation{Instituto de Ciencias Nucleares, Universidad Nacional
  Aut\'onoma de M\'exico, Circuito Exterior C.U., A.P. 70-543,
  M\'exico D.F. 04510, M\'exico}

\author{Olivier Sarbach}
\email[]{sarbach@ifm.umich.mx}
\affiliation{Instituto de F\'isica y Matem\'aticas, Universidad Michoacana de 
San Nicol\'as de Hidalgo,
Edificio C-3, Ciudad Universitaria, 58040 Morelia, Michoac\'an, M\'exico}

\date{\today}

\bigskip

\begin{abstract}
We present new, asymptotically flat, static, spherically symmetric and traversable wormhole solutions 
in General Relativity which are supported by a family of ghost scalar fields with quartic potential. This family consists of a particular composition of the scalar field modes, in which each mode is characterized by the same value of the angular momentum number $\ell$, yet the composition yields a spherically symmetric stress-energy-momentum and metric tensor. For $\ell = 0$ our solutions reduce to wormhole configurations which had been reported previously in the literature. We discuss the effects of the new parameter $\ell$ on the wormhole geometry including the motion of free-falling test particles.
\end{abstract}



\maketitle

\section{Introduction}
\label{sec:introduction}

\begin{figure}[!ht]
\includegraphics[width=0.45\columnwidth]{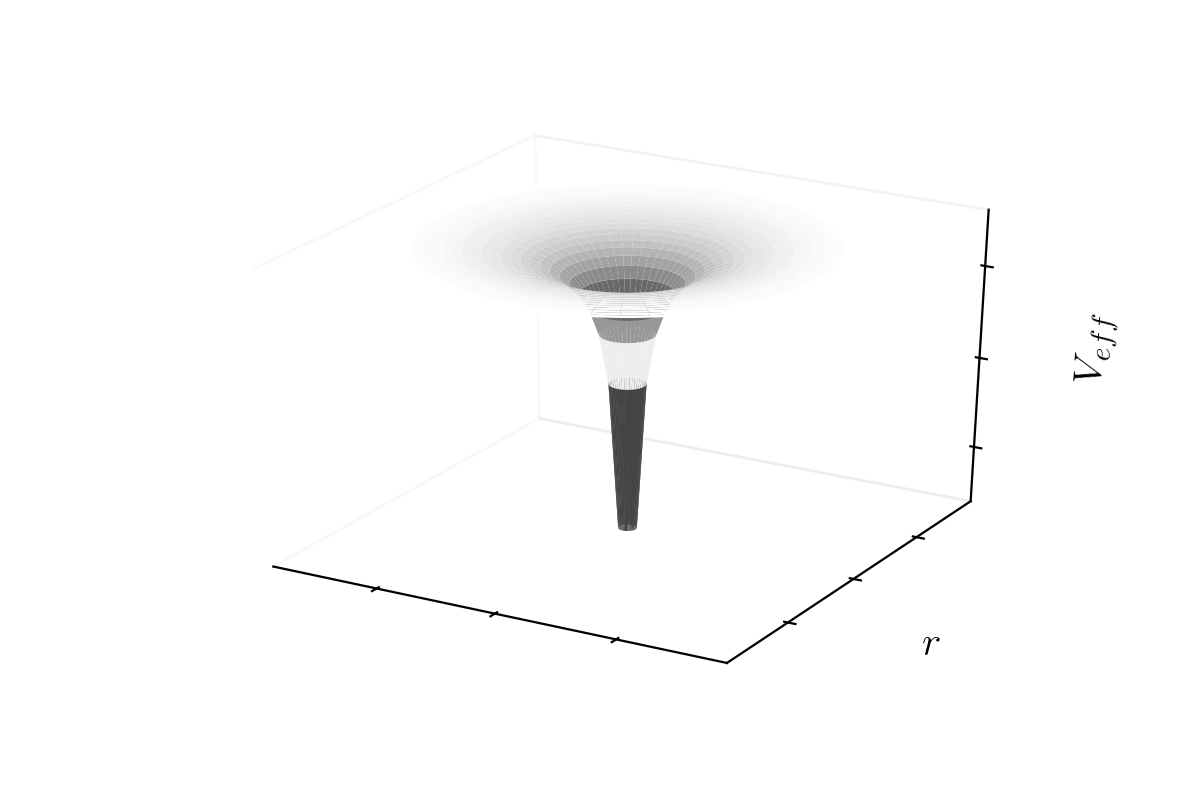}\hspace{1cm}
\includegraphics[width=0.45\columnwidth]{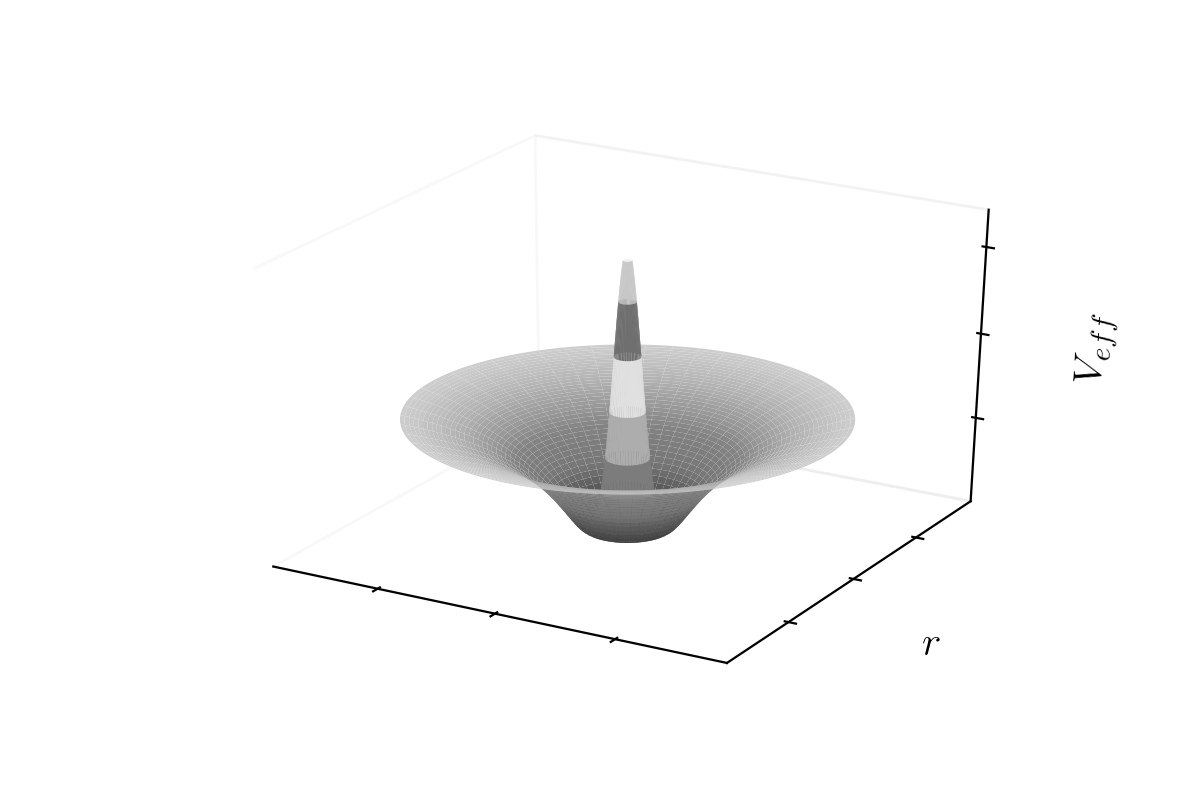}
\caption{Gravitational potential produced by a point mass $\sim\,-1/r$ (left panel) 
and by the exotic distribution presented in this work (right panel), see section \ref{sec:geodesics}. 
In both cases a test particle with vanishing angular momentum is considered.}
 \label{fig:pots}
\end{figure}

The essential property of General Relativity, namely, that matter determines the geometry of the 
spacetime, acquires a new light when the matter is such that it violates the 
energy conditions~\cite{Krasnikov:1999ie,Visser:2003yf,Lobo:2004rp,Lobo:2005us,
Lobo:2005yv}. In particular, the violation of the null energy condition opens the possibility for the existence of globally hyperbolic, asymptotically flat spacetimes with non-trivial topological structures~\cite{Friedman:1993ty}. Such matter, usually referred to as {\it exotic} in the literature, generates peculiar responses in the properties of the spacetime curvature with important consequences on the effective gravitational potential, producing potential ``bumps" instead of the usual potential wells. To provide an explicit example, in Fig.~\ref{fig:pots}, we present the gravitational effective potentials for a massive, radially infalling test particle for two cases: the first one is due to the presence of a point mass which generates the usual gravitational well, whereas the second one is generated by a distribution of exotic matter (as the one discussed later in this work) in which case the potential exhibits a different type of convexity corresponding to a gravitational potential bump.

Moreover, under specific circumstances, the {\it bump} could be such that it connects two separated regions of spacetime. The resulting configuration is dubbed {\it wormhole}, and it offers challenges and opportunities to better understand the relation between matter and geometry, aside from the fact that, being bona fide solutions to Einstein's equations, it could potentially describe an astrophysical scenario if exotic matter turns out to actually being present in our Universe.

In cosmology, matter with negative pressure can be used to describe the observed accelerated expansion of the Universe~\cite{ArmendarizPicon:2000dh,Garriga:2000cv,Huterer:2000mj, Copeland:2006wr} and seems to be favored by several observational constraints~ 
\cite{EscamillaRivera:2011qb,Kazin:2014qga,Ade:2013zuv}. Additionally, modeling the dark energy with an equation of state of the form $p=\omega\rho$, the observations suggest a value of $\omega$ close to $-1$ or even smaller, in which case the existence of astrophysical or cosmological wormholes becomes plausible.

The studies of traversable wormholes have their origin with Ellis' work~\cite{Ellis:1973yv},  where the author presented a black hole like solution to Einstein's equations, and in order to remove the singularity, used a scalar field and  {\it drain the hole}. Actually, the same solution, based on a different approach was obtained almost at the same time by Bronnikov~\cite{Bronnikov:1973fh}. 
It turned out that the solution represented a bridge between two regions of the spacetime~\cite{Abe:2010ap}. Over the years, the idea was further developed, and the best known example of a traversable wormhole appeared in 1988, in the work of Morris and Thorne~\cite{Morris:1988cz}. Since then, a plethora of literature has arisen and the complexity of the models has increased, see for example~\cite{Lobo:2007zb, Visser:1995cc} and references therein.

In order to obtain a wormhole solution to the Einstein equations, some works use generation
procedures, such as the Newman-Janis algorithm, which allows to obtain a 
Kerr black hole solution starting from a Schwarzschild one; in this way, a rotating (although not asymptotically flat) solution was obtained starting from one of the original Ellis models \cite{Matos:2005uh}. There is also a technique which uses the thin shell approach, which assumes that the matter is concentrated in a three-dimensional submanifold. However, a common practice is to analyze the geometry describing a putative wormhole without mentioning the possible matter that could generate it, artificially producing general forms of such wormholes which might even include rotation~\cite{Teo:1998dp}; in the words of Morris and Thorne: fixed the geometry... {\it ``and let the builders of a wormhole synthesize, or search throughout the universe for, materials or fields with whatever stress-energy tensor might be required''} \cite{Morris:1988cz}.

In the present work, we prefer to avoid this ``reversed engineering approach'' and 
assume the specification of a suitable matter model which allows for a large class of static, spherically symmetric and traversable wormhole solutions. In particular, following the recent approach in~\cite{Alcubierre:2018ahf} to construct a generalized class of static and spherically symmetric boson stars, we consider a family of massive, complex and self-interacting ghost scalar fields similar to the one considered in Dzhunushaliev \textit{et al.}~\cite{Dzhunushaliev:2017syc, Dzhunushaliev:2008bq}, but which includes an extra parameter $\ell$ mimicking the effects of the angular momentum. In this way, new spherical and traversable wormhole solutions can be constructed which generalize those of Refs.~\cite{Dzhunushaliev:2017syc, Dzhunushaliev:2008bq} to $\ell > 0$. Accordingly, and following the terminology of the $\ell$-boson stars, we dub these solutions $\ell$-wormholes.

While Ellis' original solutions use a massless, time-independent real scalar field without self-interaction, in this work we consider massive, complex and self-interacting scalar fields with a harmonic time-dependency \textit{\`a la} Dzhunushaliev \textit{et al.}~\cite{Dzhunushaliev:2017syc, Dzhunushaliev:2008bq}, but instead of considering just a single field we consider a family of fields with angular momentum number $\ell m$ with $\ell$ fixed and $m=-\ell,\ldots,\ell$. Assuming like in~\cite{Alcubierre:2018ahf} that each of these fields has exactly the same radial dependency, we obtain static, spherically symmetric wormhole solutions. When $\ell = 0$, the mass of the scalar field, its self-interaction and the time-frequency vanish one recovers Ellis' wormhole solutions.

The new wormholes have several interesting characteristics, such as curvature scalars and effective potentials which smooth out the features of the corresponding $0$-wormholes. The geodesic motion helps us to understand the role played by the $\ell$--parameter in the spacetime configuration. Finally, the presence of a new parameter gives rise to the possibility that this wormhole might be stable, a feature that will be discussed in a followup work.

The paper is organized as follows. In Section~\ref{sec:base}, we specify our metric ansatz describing static, spherical symmetric and traversable wormhole spacetimes and introduce the matter model. 
In Section~\ref{sec:wormhole}, we derive the static field equations in spherically symmetry, discuss some qualitative properties of the wormhole solutions and then construct numerical solutions to the field equations whose main properties are discussed next in Section~\ref{sec:properties}. In Section~\ref{sec:nembedding}, we discuss the embedding diagrams visualizing the spatial geometry of the solutions, derive the geodesic equations for massive or massless test particles propagating in the wormhole metric and analyze the motion under several conditions determined by the wormhole parameters. Finally, we discuss and summarize our results in Section~\ref{sec:Conclusions}.

\section{Foundations}
\label{sec:base}
The determination of the stress-energy-momentum tensor that supports a wormhole geometry is of the utmost importance to understand its physical properties and structure. As already mentioned in the introduction, an asymptotically flat wormhole geometry in general relativity requires the matter to be {\it exotic}, that is, matter that does not fulfill the regular properties of the usual matter we deal with everyday.\footnote{However, it should be mentioned that there are examples of traversable wormholes without exotic matter in modified theories of gravity~\cite{Kanti:2011jz} or in general relativity when the asymptotic flatness condition is replaced by adS-asymptotics~\cite{Ayon-Beato:2015eca}.} More specifically, the matter must violate the null energy condition, ${{T}}_{\mu\nu}k^{\mu}k^{\nu} \geq 0$, where ${{T}}_{\mu\nu}$ is the stress-energy-momentum tensor and $k^{\mu}$ any null vector \cite{Morris:1988cz,Visser:1995cc,Visser:1999de}. Incidentally, this is also the fundamental ingredient of the so-called ghost energy, a model not 
excluded by observations to be a candidate for dark energy. For instance, 
constraints from the Supernovae Ia Hubble diagram \cite{Majerotto:2004ji} favor 
the existence of an equation of state for such dark fluid, $p = \omega\rho$ with $\omega<-1$, a model consistent with ghost energy~\cite{Lobo:2005us}.

In practice, violation of the null energy condition is accomplished by changing the global sign in the stress-energy-momentum tensor in Einstein's equations. Ellis called this the {\it other polarity of the equations} \cite{Ellis:1973yv}. This change in sign in the equations is attributed to the 
type of matter, and has multiple implications which might lead to misunderstandings. 
A  global change in sign to the stress-energy-momentum tensor implies that the usual 
definition of density also has the opposite sign and is thus negative.

In this work we interpret the physical properties of the wormhole directly in terms of the theory of general relativity and Einstein's field equations, so that the exotic matter produces a different reaction in the curvature of the spacetime, particularly in the effective potential in which the test particles move, generating bumps instead of wells, so that a particle has to spend potential energy in order to get closer to the source, while it gains kinetic energy and accelerates when getting away from it; like when climbing a mountain to reach the summit and then going down.

In particular, we stress that when talking about test particles we assume the validity of the weak equivalence principle, which assumes that the inertial and gravitational masses are equal to each other. Therefore, free-falling test particles or photons always follow causal geodesics of the underlying spacetime, regardless of the sign of their mass.\footnote{However, see~\cite{doi:10.1119/1.17293} for bizarre implications in systems involving hypothetical point particles with positive and negative masses.} The ``bump interpretation" mentioned so far will become evident when analyzing the geodesic motion of test particles in Section~\ref{sec:nembedding}.

\subsection{Metric ansatz}

We will consider a static spherically symmetric spacetime with a line element of the form:
\begin{equation}\label{eq:metric}
ds^2=-a(r)\,c^2\,dt^2 + a(r)^{-1}\,dr^2 +
R^2(r)\,d\Omega^2, 
\end{equation} 
where $R$ and $a$ are positive functions only of the 
radial coordinate $r$, and $d\Omega^2=d\theta^2 + \sin^2\theta\,d\varphi^2$.
Notice that for $R^2=r^2 + b^2$, with $b$ a positive constant, and $a=1$, the reflection-symmetric Ellis wormhole metric is recovered~\cite{Ellis:1973yv} and, from it, with a suitable redefinition of the radial coordinate, one obtains the usual form of the Morris--Thorne like wormhole~\cite{Morris:1988cz}. Also note that the coordinate $r$ our work is based on extends from $-\infty$ to $+\infty$, and we will demand that $R$ be regular at the throat $r = 0$, which corresponds to a minimum of the area $4\pi R^2$ of the invariant two-spheres.

\subsection{Matter content}

In the present work, we consider a set of several massive scalar fields with a 
self-interaction term. Our configurations are constructed in such a way that the sum of the fields 
preserves the spherical symmetry of the stress-energy-momentum tensor and includes an 
extra parameter associated with the angular momentum number 
$\ell$. This approach was introduced in~\cite{Olabarrieta:2007di} in the context of critical collapse, and recently used in~\cite{Alcubierre:2018ahf} to construct $\ell$-boson stars.

We start with the Lagrangian density for $N$ complex massive scalar fields

\begin{equation}
\mathcal{L}_{\Phi}= 
-\frac{1}{2\kappa}\left[\sum_{i=1}^N\eta\nabla_\mu\Phi_i\nabla^\mu\Phi_i^*+V(\ 
\abs{\Phi}^2)\right],
\end{equation}
with a quartic potential
\begin{equation}
\begin{split}
V(\abs{\Phi}^2) &=\sum_{i=1}^N V^{(i)}= \sum_{i=1}^N\left(\eta_\mu\frac{ 
m_{\Phi}^2c^2}{\hbar^2}|\Phi_i|^2+\eta_\lambda\frac{\lambda}{2\hbar^2}\abs{
\Phi_i}^2\sum_{j=1}^N\abs{\Phi_j}^2\right)
\end{split}
\end{equation}
where 
$\kappa=8\pi G/c^4$, $\hbar$ is the 
reduced Planck constant, $m_{\Phi}$ is the mass of the scalar field particle and 
$\lambda$ is the parameter measuring the strength of the quartic interaction 
term.
The values $\eta=\eta_\mu=\eta_\lambda=1$ represent the canonical scalar 
fields while $\eta=\eta_\mu=-\eta_\lambda=-1$ describe the type of ghost fields 
in 
which we will be interested in, and from now on we fix the latter choice. 
In the following, for convenience, we will work with the rescaled quantities 
$\mu=m_\Phi c/\hbar$ and $\Lambda=\lambda/2\hbar^2$ instead of $m_\Phi$ and 
$\lambda$.

The stress-energy-momentum tensor associated with the scalar field $\Phi_i$ is thus given by 
\begin{equation}
T^{(i)}_{\mu\nu}=\frac{c^4}{16\pi G}\,
\left[ -\left(\nabla_{\nu}{\Phi_i}\,\nabla_{\mu}{\Phi^*_i} +
\nabla_{\nu}{\Phi_i}\,\nabla_{\mu}{\Phi^*_i}\right) -
g_{\mu\nu}\left( -
\nabla_{\alpha}{\Phi_i}\nabla^{\alpha}{\Phi^*_i}
+ V^{(i)}\right)
\right],
\label{eq:stressenergy1f}
\end{equation}
while the total stress-energy-momentum tensor that we plug into Einstein's 
field equations is
\begin{equation}
T_{\mu\nu}=\sum_{i=1}^N T^{(i)}_{\mu\nu} .\label{eq:stressenergyfull}
\end{equation}
In~\cite{Olabarrieta:2007di}, for the case of real scalar fields, and in the appendix of~\cite{Alcubierre:2018ahf}, for complex ones, it was shown that for an appropriate superposition, a stress-energy-momentum tensor of the form~(\ref{eq:stressenergyfull}) with $\Lambda=0$ may be spherically symmetric, even though the individual fields $\Phi_i$ have non-vanishing angular momentum. Here, we generalize this result further to include the self-interaction of the field. The procedure is as follows.

Each scalar field $\Phi_i$ has the form
\begin{equation}\label{eq:fieldansatz}
\Phi_{i}(t,r,\theta,\varphi)=\, \phi_\ell(t, r)Y^{\ell 
	m}(\theta,\varphi),\qquad
i = \ell +1 + m,
\end{equation}
where $m$ varies over $-\ell,-\ell+1,\ldots,\ell$ (such that $i$ varies from $1$ to $N = 2\ell + 1$) and $Y^{\ell m}(\theta,\varphi)$ denote the standard spherical harmonics. Here, the parameter $\ell$ is kept fixed and the amplitudes $\phi_\ell(t, r)$ are equal to each other for all $m$. Using the addition theorem for the spherical harmonics~\cite{Jackson:1998nia}, one can show that the resulting stress-energy-momentum tensor in Eq.~(\ref{eq:stressenergyfull}) for the $N = 2\ell+1$ fields is spherically symmetric.

Next, one considers a stationary state with harmonic time dependence for the 
scalar field
\begin{equation}
\phi_\ell(t,r)=e^{i\omega t}\,\chi_\ell(r), 
\end{equation}
where $\chi_\ell$ is function of $r$ and $\omega$ is a real 
constant. Once such procedure is carried out, the following non-trivial components of the stress-energy-momentum tensor are obtained:
\begin{eqnarray}
{T^t}_t&=&\frac{c^4}{8 \pi G}\frac{2 \ell+1}{8 \pi} 
\left\{a \left(\frac{d \chi_\ell}{d r}\right)^2
+ \left[\frac{\ell \left(\ell+1\right)}{R^2} + 
\mu^2 - 
\frac{2\ell+1}{4\pi}\Lambda\chi_\ell^2+ 
\frac{1}{a} \omega^2\right]
{\chi_\ell}^2\right\}, \label{eq:Tttf}\\
{T^r}_r&=&\frac{c^4}{8 \pi G} \frac{2 \ell+1}{8 \pi} 
\left\{ -a \left(\frac{d \chi_\ell}{d r}\right)^2 + 
\left[\frac{\ell \left(\ell+1\right)}{R^2} + 
\mu^2 -\frac{2\ell+1}{4\pi}\Lambda \chi_\ell^2- 
\frac{1}{a} \omega^2\right]
{\chi_\ell}^2\right\}, \label{eq:Trrf}\\
{T^\theta}_\theta&=&{T^\varphi}_\varphi=
\frac{c^4}{8 \pi G}  \frac{2 \ell+1}{8 \pi} 
\left\{a \left(\frac{d \chi_\ell}{d r}\right)^2 +  
\left[\mu^2 -\frac{2\ell+1}{4\pi}\Lambda \chi_\ell^2 - 
\frac{1}{a} \omega^2\right]
{\chi_\ell}^2\right\}. \label{eq:Tththf}
\end{eqnarray}

Notice how the procedure of adding individual stress-energy-momentum tensors maintains 
the spherical symmetry and yields a result that depends on the angular momentum number $\ell$ through the centrifugal-like terms $\ell(\ell+1)/R^2$. As expected and shown below, this dependency plays a nontrivial role in the solutions of Einstein's equations. The mixed components ${T^t}_r$, ${T^t}_\theta$ and ${T^t}_\varphi$ vanish; indicating 
that there are no fluxes of matter in this case, which is compatible with the assumption of staticity of the metric.

Notice also that the stress-energy-momentum tensor~(\ref{eq:Tttf}--\ref{eq:Tththf}) violates the null energy condition everywhere; for instance, the null vector field $k = a^{-1/2} c^{-1}\partial_t + a^{1/2}\partial_r$ gives
\begin{equation}
T_{\mu\nu} k^\mu k^\nu = -T^t{}_t + T^r{}_r
 = -\frac{c^4}{8 \pi G}  \frac{2 \ell+1}{4\pi} \left[ a\left(\frac{d \chi_\ell}{d r}\right)^2
 + \frac{\omega^2}{a} \chi_\ell^2 \right],
\end{equation}
which is negative unless the scalar field vanishes.

Now, one can compute the equation of motion for each individual field, {\it i. e.} the Klein-Gordon equation, using the fact that the divergence of the total stress-energy-momentum tensor is zero.
Each amplitude obeys the identical equation:
\begin{equation}
\frac{d}{d\,r}\left[ a
\,R^2
\frac{d\chi_\ell}{d r}\right] + 
 R^2\,\left(\frac{\omega^2}{a} 
- \frac{\ell \left(\ell+1\right)}{R^2}
 -\mu^2 + 
\frac{2\ell+1}{2\pi}\Lambda \chi_\ell^2 \right) \chi_\ell=0,
\label{eq:KG}
\end{equation}
where we have used the fact that spherical harmonics are eigenfunctions
of the Laplace-Beltrami operator
\begin{equation}
\Delta_{S^2}\,Y^{\ell m}=
\left(\frac{\partial^2}{\partial\theta^2} + 
\cot\theta\frac{\partial}{\partial\theta} 
+ \frac{1}{\sin^2\theta}\frac{\partial^2}{\partial\varphi^2}
\right)Y^{\ell m}=-\ell\left(\ell+1\right)Y^{\ell m} \ . 
\end{equation}

\begin{figure}[!ht]
\includegraphics[width=0.7\columnwidth]{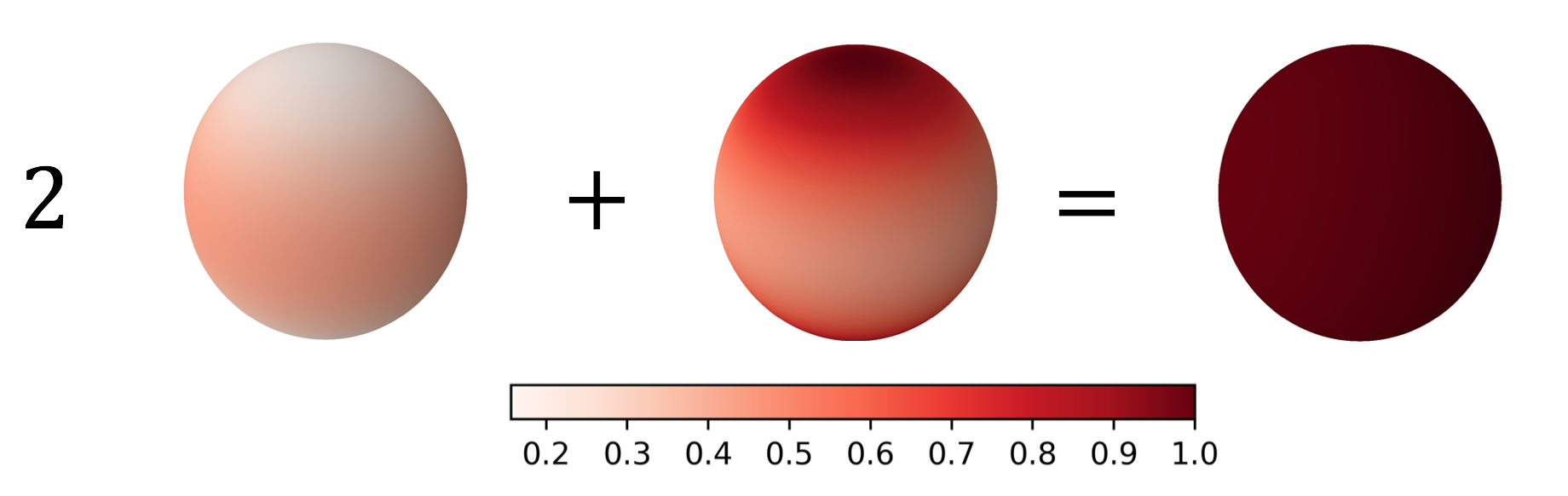}
 \caption{Normalized density at the throat for a $\ell=1$ wormhole. 
 The first sphere represents the $m=\pm1$ contribution while the center sphere 
corresponds to the $m=0$ contribution.}
 \label{fig:harmonics}
\end{figure}

As an example of the construction of the $\ell-$wormhole by the contribution 
of individual non-spherical scalar fields, in Fig.~\ref{fig:harmonics} we show 
the distribution of the density at the throat for the fields 
$(\phi_1Y^{1-1},\phi_1Y^{10},\phi_1Y^{11})$. The values are given by 
Eq.~(\ref{eq:stressenergyfull}) for the ${T^t}_t\propto\rho$ component. This is 
the case for $\ell = 1$ wormhole, so that there are three values for $m$. 
The first sphere represents the sum of the $m = -1$ and $m = 1$ contributions, 
the second one represents the $m=0$ field, and the combination is given in such a way 
that the total density (the third sphere) is spherically symmetric.

\section{Stationary Wormhole equations}
\label{sec:wormhole}

In order to obtain the remaining field equations, it is helpful to notice that
with the stress-energy-momentum tensor components given by Eqs.~(\ref{eq:Tttf},~\ref{eq:Trrf},~\ref{eq:Tththf}) the following equation is satisfied
\begin{equation}
\frac{{T^t}_t - {T^r}_r}{2} - 
{T^\theta}_\theta=-\frac{c^4}{8 \pi G}  \frac{2 \ell+1}{8 \pi} 
\left(\mu^2 -\frac{2\ell+1}{4\pi}\Lambda \chi_\ell^2-  
\frac{2}{a} \omega^2\right) \chi_\ell^2.
\end{equation}
On the other hand, from the 
line element Eq.~(\ref{eq:metric}), we obtain the Bianchi-Einstein tensor, 
${G^\mu}_\nu$, and the same linear combination of the 
components gives:
\begin{equation}
\frac{{G^t}_t - {G^r}_r}{2} - {G^\theta}_\theta=-
\frac{1}{2\,R^2} 
\frac{d}{dr}\left[R^2\, 
\frac{da}{dr}\right].
\end{equation}
Thus, with the aid of Einstein's equations:
\begin{equation}
{G^\mu}_\nu =\frac{8\,\pi\,G}{c^4}\,{T^\mu}_\nu, \label{eqs:Eins}
\end{equation}
we obtain our next field equation:
\begin{equation}
\frac{1}{2\,R^2} 
\frac{d}{dr}\left[R^2\,\frac{da}{dr}\right] =
 \frac{2 \ell+1}{8 \pi} \left(\mu^2 -\frac{2\ell+1}{4\pi}\Lambda \chi_\ell^2 - 
 \frac{2}{a} \omega^2\right) \chi_\ell^2.
\label{eq:ddaBis}
\end{equation}
Notice that when a static, massless scalar field (regular or exotic) without interaction is considered, then a particular solution is obtained in which the metric function $a$ is constant. 

A further field equation comes from the combination of the 
${{G}^t}_t$ component plus the ${{G}^r}_r$ one, and the corresponding ${T^\mu}_\nu$ components:
\begin{equation}
\frac{d\,R^2}{dr}\frac{d\,a}{dr} + a\frac{d^2\,R^2}{dr^2}-2=
\frac{2\ell+1}{4\pi}\left[R^2\,\left(\mu^2-\frac{2\ell+1}{4\pi}\Lambda\chi_\ell^2\right)+\ell(\ell+1)\right]\chi_\ell^2.
\label{eq:ddR2}
\end{equation}
As mentioned above, for the first independent field equation, we consider the 
Klein-Gordon equation, Eq.~(\ref{eq:KG}). In this way we obtain a system of equations in which each 
function $\chi_\ell(r)$, $a(r)$ and $R^2(r)$ appears as the only second 
derivative:
\begin{eqnarray}
\label{eq:ddchi}
\chi_\ell''&=&-\left(\frac{{R^2}'}{R^2} + \frac{a'}{a}\right)\chi_\ell'+ \frac{1}{
 a}\left[\mu^2 - \frac{\omega^2}{a} + \frac{\ell(\ell+1)}{R^2} - \frac{2\ell+1}{2\,\pi}\Lambda\chi_\ell^2\right]\chi_{\ell},\\
\label{eq:dda}
 a''&=&-\frac{{R^2}'}{R^2}\,a'+ \frac{2\ell+1}{4\pi}\left(\mu^2 -\frac{2}{a}\,\omega^2 - \frac{2\ell+1}{4\pi}
\Lambda\chi_\ell^2\right)\chi_\ell^2,\\
\label{eq:ddf}
{R^2}''&=&\frac{1}{a}\left\{ - a'\,{R^2}' + 2 +\frac{2\ell+1}{4\pi}\,\left[ \left(\mu^2-\frac{2\ell+1}{4\pi}
\Lambda\chi_\ell^2\right)\,R^2+ \ell\,(\ell+1)\right]\chi_\ell^2 \right\},
\end{eqnarray}
where a prime denotes derivative with respect to $r$. The remaining field equation is the $rr$-component of Eq.~(\ref{eqs:Eins}) which yields
\begin{equation}
 \frac{{R^2}'}{2\,R^2}\,\left(a' + \frac{a\,{R^2}'}{2\,R^2}\right) - \frac{1}{R^2} 
 = \frac{2\,\ell+1}{8\,\pi}
 \left[ -a\,{{\chi_\ell}^2}' + \left(\mu^2 - \frac{2\,\ell+1}{4\,\pi}\,\Lambda\,{\chi_\ell}^2 - \frac{\omega^2}{a}
 + \frac{\ell\,\left(\ell + 1\right)}{R^2}\right)\,{\chi_\ell}^2\right], 
\label{eq:cons}
\end{equation}
which can be interpreted as a constraint since it only involves zeroth and first-order derivatives of the fields. Provided the second-order field equations~(\ref{eq:ddchi},~\ref{eq:dda},~\ref{eq:ddf}) are satisfied, the twice contracted Bianchi identity $\nabla_\mu G^\mu{}_r = 0$ and $\nabla_\mu T^\mu{}_r = 0$ imply that
\begin{equation}
\frac{d}{dr} \left[ a R^4\left( G^r{}_r - \frac{8\pi G}{c^4} T^r{}_r \right) \right] = 0,
\end{equation}
such that it is sufficient to solve Eq.~(\ref{eq:cons}) at one point (the throat, say).

\begin{figure}[!ht]
	\includegraphics[width=0.45\columnwidth]{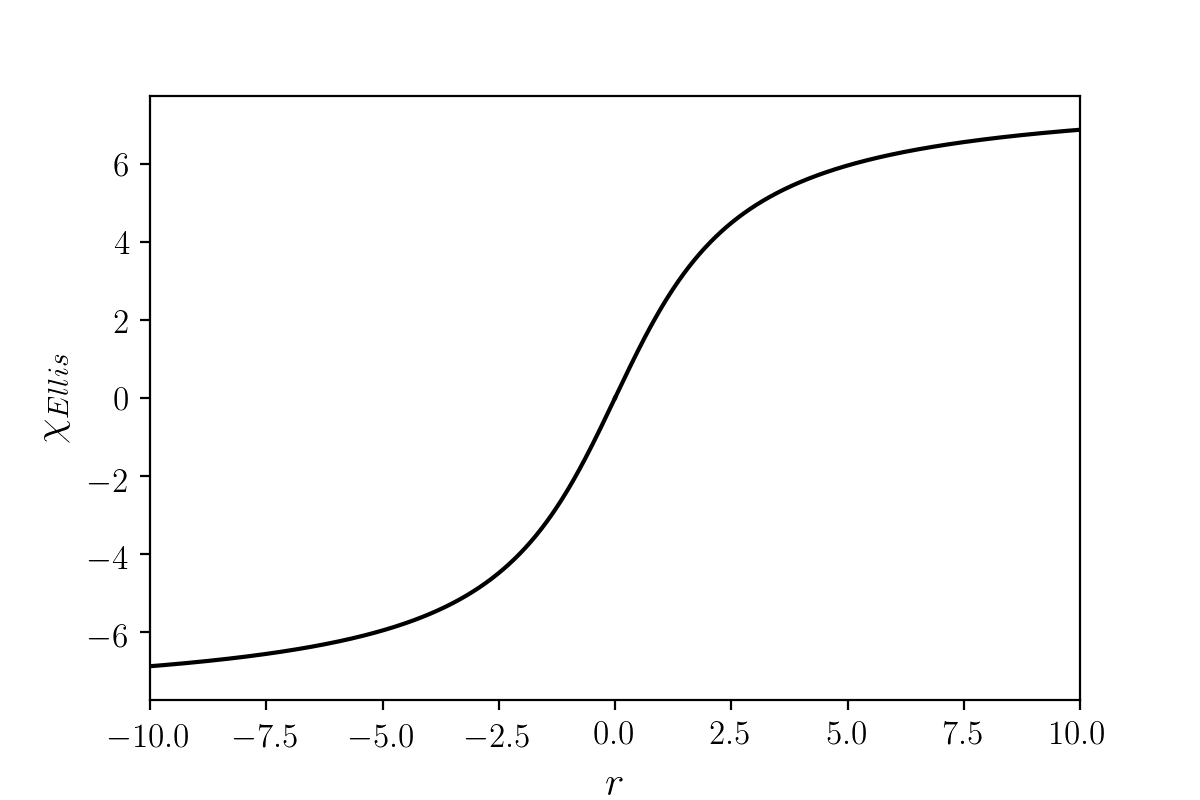}\hspace{0.5cm}
	\includegraphics[width=0.45\columnwidth]{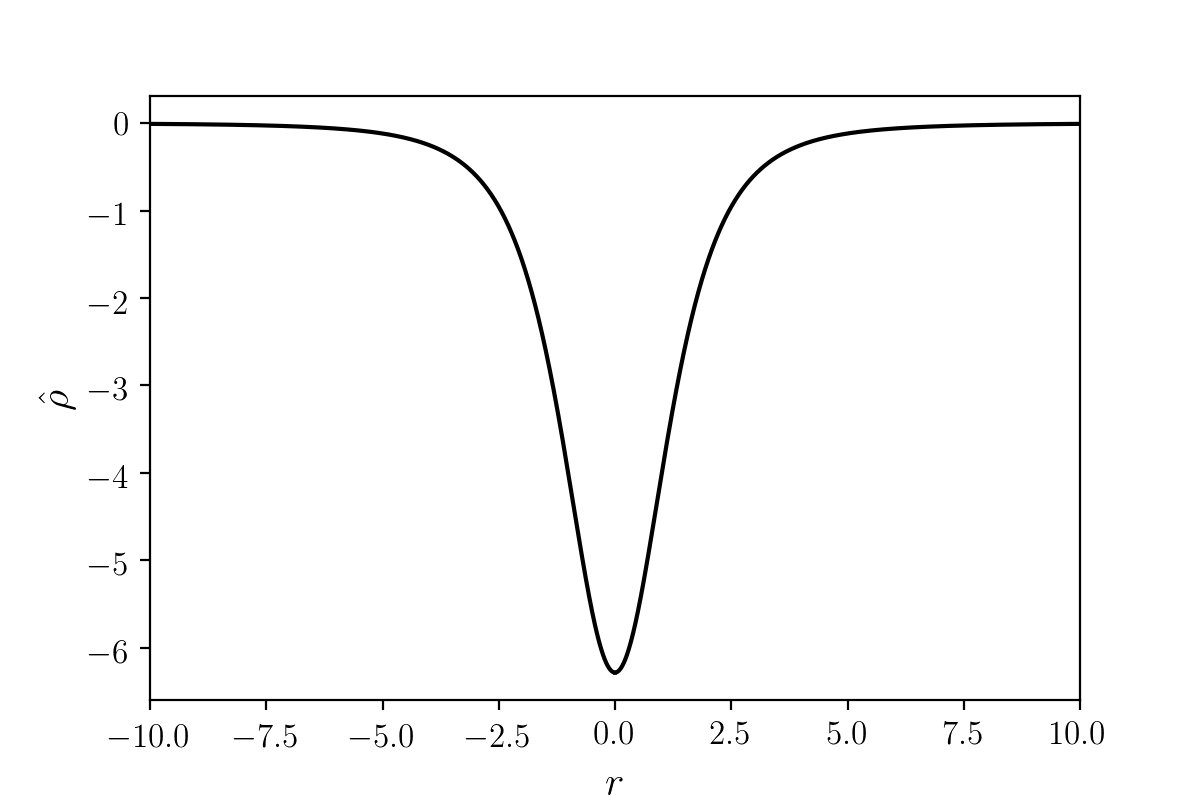}
	\caption{Ghost field and the corresponding density for the original Ellis wormhole.}
	\label{fig:chi_Ellis}
\end{figure}

\begin{figure}[!ht]
	\includegraphics[width=0.45\columnwidth]{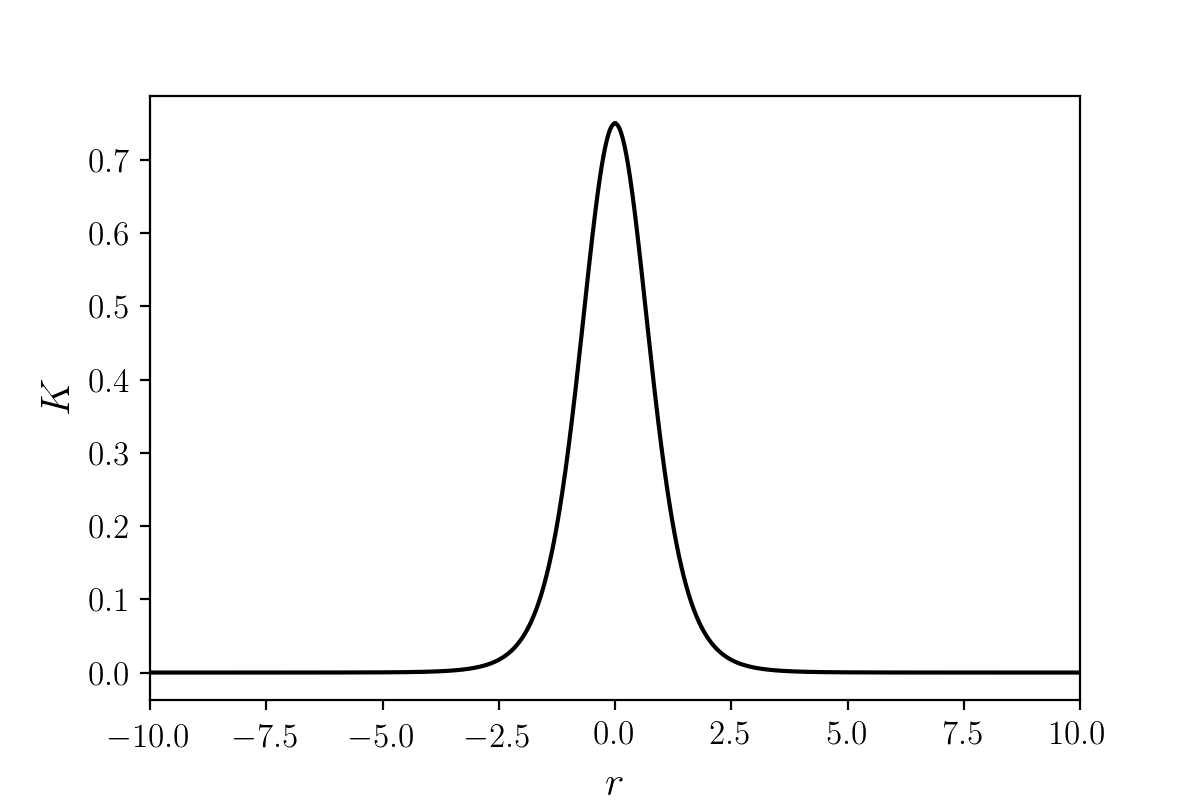}\hspace{0.5cm}
	\includegraphics[width=0.45\columnwidth]{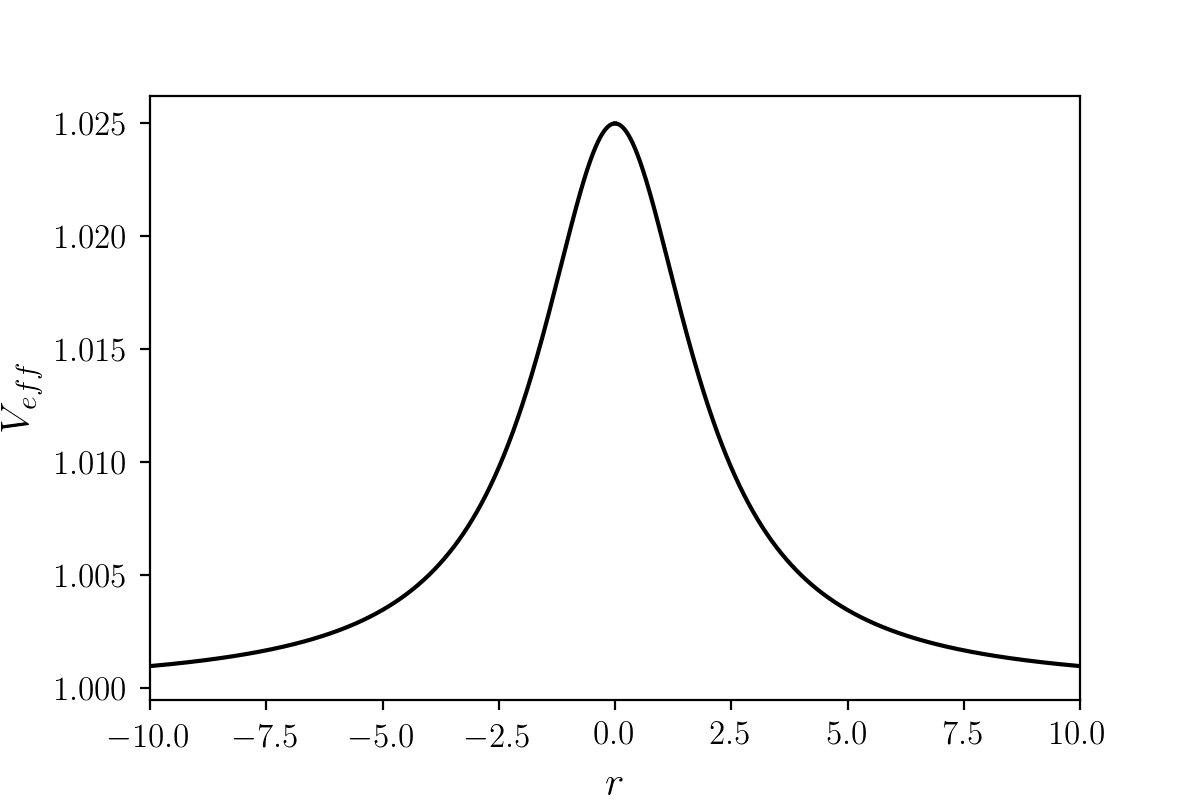}
	\caption{Kretschmann scalar  and effective potential, $V_{eff}=L^2/R^2+\kappa$ for a massive particle propagating on the reflection-symmetric Ellis wormhole.}
	\label{fig:KV_Ellis}
\end{figure}

A particular simple solution arises when a static, spherical, massless scalar field (regular or exotic) without interaction is considered. In this case, the parameters $\omega$, $\mu$, and $\Lambda$ vanish and considering $\ell=0$, the field equations~(\ref{eq:ddchi}--\ref{eq:cons}) can be integrated explicitly~\cite{Ellis:1973yv,Bronnikov:1973fh}, see also~\cite{Gonzalez:2008wd}. The simplest (but not unique) solution is obtained assuming that the metric function $a$ is constant. This yields the solution
\begin{equation}
a = 1,\qquad
R^2 = b^2 + r^2,\qquad
\chi_{\text{Ellis}}\left(r\right)=\sqrt{8\pi}\arctan\left(\frac{r}{b}\right),
\label{eqs:Ellis_sol}
\end{equation}
which has the property that the metric functions $a$, $R^2$ and the gradient of 
$\chi_{\text{Ellis}}$ are reflection symmetric about the throat $r = 0$. In 
Fig.~\ref{fig:chi_Ellis}, we present the plot of Ellis' 
ghost field and the corresponding energy density. 
Although the scalar field 
itself does not decay to zero simultaneously at both asymptotic ends $r\to 
\pm\infty$, its gradient does. Since in the massless case the stress-energy-momentum tensor
and equations of motion only depend on the gradient of the scalar field, the configuration is 
localized from a physical point of view. Furthermore, we observe that the 
density is negative everywhere. The curvature and Kretschmann scalars are given 
by $R_{s,{\rm Ellis}}=-\frac{2\,b^2}{\left(r^2+b^2\right)^2}$ and $K_{\rm 
Ellis}= R_{s,{\rm Ellis}}^2$ (see Fig.~\ref{fig:KV_Ellis}), respectively, and like the density, they have a 
fixed sign.

In the following, we consider much more general wormhole solutions in which the parameters $\omega$, $\ell$, $\mu$ and $\Lambda$ do not necessarily vanish. These solutions are obtained by numerically integrating the field equations~(\ref{eq:ddchi},~\ref{eq:dda},~\ref{eq:ddf}) and taking into account the constraint~(\ref{eq:cons}). For simplicity, in this article, we restrict ourselves to the reflection-symmetric case (although more general wormhole solutions which are asymmetric about the throat could also be considered). These solutions satisfy the following boundary conditions at the throat, $r = 0$:
\begin{eqnarray}
\chi_\ell'(0)=0,
\label{eq:dchi}\\
a'(0)=0,
\label{eq:da}\\
{R^2}'(0)=0.
\label{eq:dR2}
\end{eqnarray}
Denoting by $b := R(0)$ the areal radius of the throat, the constraint~(\ref{eq:cons}) yields the following condition at $r = 0$:
\begin{equation}
\left[ 1 + \frac{(2\ell+1)\ell(\ell+1)}{8\pi}\chi_\ell(0)^2 \right] b^{-2}
 = \frac{2\ell+1}{8\pi}\left( \frac{\omega^2}{a(0)} - \mu^2 + \frac{2\ell+1}{4\pi}\Lambda \chi_\ell(0)^2
 \right)\chi_\ell(0)^2,
\label{eq:throat_cons}
\end{equation}
which fixes the radius $b$ of the throat and requires $a(0)$ and $\chi_\ell(0)\neq 0$ to be chosen such that
\begin{equation}
\frac{\omega^2}{a(0)} + \frac{2\ell+1}{4\pi}\Lambda \chi_\ell(0)^2 > \mu^2.
\label{eq:throat_ineq}
\end{equation}
Note that this inequality and Eq.~(\ref{eq:dda}) also imply that $a$ has a 
local maximum at the throat. Next, Einstein's equation~(\ref{eq:ddf}) together 
with the conditions~(\ref{eq:dchi},~\ref{eq:da},~\ref{eq:dR2}), implies the 
relation
\begin{equation}
\frac{1}{2} a(0) (R^2)''(0) =\frac{2\ell+1}{8\pi}\left[b^2\left(\mu^2-\frac{2\ell+1}{4\pi}
\Lambda\chi_\ell(0)^2\right)+\ell(\ell+1)\right]\chi^2_\ell(0)+1.
\label{eq:reg}
\end{equation}
Using Eq.~(\ref{eq:throat_cons}) this can be simplified considerably,
\begin{equation}
\frac{1}{2} a(0) (R^2)''(0) = \frac{2\ell+1}{8\pi} \frac{b^2\omega^2}{a(0)}\chi_\ell(0)^2,
\label{eq:throat_dd}
\end{equation}
which shows that the throat is indeed a local minimum\footnote{For $\omega\neq 
0$ 
the right-hand side of Eq.~(\ref{eq:throat_dd}) is positive since $\chi_\ell$ 
and $\chi_\ell'$ cannot both vanish at $r = 0$; otherwise it would follow from 
Eq.~(\ref{eq:ddchi}) that $\chi_\ell$ vanishes identically. For the special case 
$\omega = 0$ see the proof of Theorem~\ref{Thm:Convex} below.} of $R^2$.
For $\ell = 0$ the Eqs.~(\ref{eq:throat_cons},~\ref{eq:throat_dd}) reduce to 
the corresponding equations in Ref.~\cite{Dzhunushaliev:2017syc} (see their equation (18) and the unnumbered equation below it.) Due to Eq.~(\ref{eq:throat_cons}), one has two free parameters at 
the throat, given by $\chi_\ell(0)\neq 0$ and $a(0)$, say. As can be checked, the 
field equations~(\ref{eq:ddchi},\ref{eq:dda},~\ref{eq:ddf},~\ref{eq:cons}) as well as 
the conditions~(\ref{eq:throat_cons},~\ref{eq:throat_dd}) are invariant with respect to the transformations
\begin{equation}
(\omega,t,r,a,R^2,\chi_\ell) 
 \mapsto \left( \frac{\omega}{\sqrt{B}},\sqrt{B}t,\frac{r}{\sqrt{B}},\frac{a}{B},R^2,\pm \chi_\ell \right),
 \label{transf}
\end{equation}
with $B > 0$ a real parameter. Therefore, one can fix the value of $a(0)$ to one, say, and adjust 
the value of $B$ such that $a(r)\to 1$ for $r\to \infty$. In this way, one is left with 
just one shooting parameter ($\chi_\ell(0) > 0$, say) at the throat $r = 0$.

At $r\to \pm\infty$, we require asymptotic flatness,
\begin{eqnarray}
\chi_\ell(r) \to 0,
\label{eq:chi_asym}\\
a(r) \to 1,
\label{eq:a_asym}\\
\frac{R(r)}{r} \to 1.
\label{eq:R_asym}
\end{eqnarray}
Under these assumptions, the field equation~(\ref{eq:ddchi}) for the scalar field reduces to
\begin{equation}
\chi_\ell''(r) \approx (\mu^2 - \omega^2)\chi_\ell(r), \label{assymp}
\end{equation}
which shows that\footnote{The limiting value $\omega^2 = \mu^2$ is discussed in Ref.~\cite{Dzhunushaliev:2017syc}.} 
\begin{equation}
\omega^2 < \mu^2
\label{eq:omega_bound}
\end{equation}
is required to have the exponentially decaying 
solution $\chi_\ell(r) \approx e^{-\sqrt{\mu^2 - \omega^2} r}$.
Approximating the (exponentially decaying) right-hand sides of Eqs.~(\ref{eq:ddaBis},~\ref{eq:ddR2}) to zero, one obtains the following behavior of the metric coefficients in the asymptotic region:
\begin{equation}\label{eq:simeq}
a \approx e^{-\frac{c_0}{r + c_1}},\qquad 
R^2\approx (r + c_1)^2 e^{\frac{c_0}{r + c_1}},\qquad
r\to \infty,
\end{equation}
for some constants $c_0$ and $c_1$.

\subsection{Qualitative analysis of the solutions}

Before numerically constructing the wormhole solutions, we make a few general remarks regarding the restrictions on the parameters $\omega$, $\mu$ and $\Lambda$ and the initial condition $\chi_\ell(0)$ and regarding the qualitative properties of the solutions. We assume in the following that 
$(\chi_\ell(r),a(r),R^2(r))$ is a smooth solution of Eqs.~(\ref{eq:KG},~\ref{eq:ddaBis},~\ref{eq:ddR2}) 
(or, equivalently, of Eqs.~(\ref{eq:ddchi},~\ref{eq:dda},~\ref{eq:ddf})) on the interval $[0,\infty)$ which 
satisfies $a > 0$, $R^2 > 0$, the boundary conditions~(\ref{eq:dchi},~\ref{eq:da},~\ref{eq:dR2}) at $r=0$ 
and~(\ref{eq:chi_asym},~\ref{eq:a_asym},~\ref{eq:R_asym}) at $r\to \infty$, and is subject to the 
conditions~(\ref{eq:throat_cons},~\ref{eq:throat_dd}) at the throat. We had already observed that an 
exponentially decaying solution at infinity requires $\omega^2 \leq \mu^2$. Furthermore, at the throat, the inequality~(\ref{eq:throat_ineq}) needs to be satisfied. A first immediate consequence of this last inequality is that the parameters $\omega$ and $\Lambda$ cannot be both zero. In fact, one has the following stronger result which shows that the self-interaction term is needed.

\begin{theorem}
There are no reflection-symmetric solutions with the above properties if $\Lambda = 0$.
\end{theorem}

\proof We prove the theorem by contradiction. If $\Lambda = 0$, the inequality~(\ref{eq:throat_ineq}) implies that
\begin{equation}
\mu^2 <\frac{\omega^2}{a(0)} \leq \frac{\mu^2}{a(0)},
\end{equation}
which requires $a(0) < 1$ and $\omega^2 > 0$. However, Eq.~(\ref{eq:dda}) with $\Lambda = 0$ implies that at 
any point $r = r_c$ where the derivative of $a$ vanishes, the equality
\begin{equation}
a''(r_c) = \frac{2\ell+1}{4\pi}\left( \mu^2 - \frac{2\omega^2}{a(r_c)} \right)\chi_\ell(r_c)^2
\label{eq:dda_rc}
\end{equation}
holds. Since $\mu^2 - 2\omega^2/a(0) < 0$, $a$ has a local maximum at the throat, as already remarked above, such that $a(r)$ decreases for $r > 0$ small enough. Since $a(r)\to 1$ as $r\to \infty$ there must be a point $r = r_c$ for which $a$ ceases to decrease, corresponding to a (local) minimum of $a$. At this point, we must have $a(r_c) < a(0)$, $a'(r_c) = 0$ 
and $a''(r_c)\geq 0$. On the other hand, since
\begin{equation}
\mu^2 - \frac{2\omega^2}{a(r_c)} < \mu^2 - \frac{2\omega^2}{a(0)} < 0,
\end{equation}
Eq.~(\ref{eq:dda_rc}) implies $a''(r_c) < 0$, provided that $\chi_\ell(r_c) \neq 0$, which leads to a 
contradiction. If $\chi_\ell(r_c) = 0$, we do not obtain an immediate contradiction since in this case it 
follows that $a''(r_c) = 0$. However, in this case, we must have $\chi_\ell'(r_c)\neq 0$ since 
otherwise $\chi_\ell$ (as a solution of the second-order equation~(\ref{eq:ddchi})) would be identically 
zero. By differentiating Eq.~(\ref{eq:dda}) twice with respect to $r$ and evaluating at $r = r_c$ one obtains $a'''(r_c) = 0$ and
\begin{equation}
a''''(r_c) = \frac{2\ell+1}{2\pi}\left( \mu^2 - \frac{2\omega^2}{a(r_c)} \right)\chi'(r_c)^2 < 0,
\end{equation}
which shows that $r = r_c$ is a local maximum of $a$ and yields again a contradiction. This concludes the proof of the theorem.
\qed

The next result implies that there cannot be more than one throat.

\begin{theorem}
\label{Thm:Convex}
Under the assumptions stated at the beginning of this subsection, the function $R^2(r)$ is strictly monotonously increasing and strictly convex on the interval $[0,\infty)$.
\end{theorem}

\proof By combining Eqs.~(\ref{eq:ddf},\ref{eq:cons}) one obtains the simple equation
\begin{equation}
\frac{(R^2)''}{R^2} = \frac{1}{2}\left[ \frac{(R^2)'}{R^2} \right]^2
 + \frac{2\ell+1}{4\pi}\left( \chi_\ell'^2 + \frac{\omega^2}{a^2}\chi_\ell^2 \right)
\label{eq:ddR2_comb}
\end{equation}
for $R^2$, which shows that $(R^2)''\geq 0$ and hence that $R^2$ is convex. We show further that the right-hand side of Eq.~(\ref{eq:ddR2_comb}) cannot vanish at any point. This is clearly the case if $\omega\neq 0$ since $\chi_\ell'$ and $\chi_\ell$ cannot vanish at the same point (otherwise it would follow from Eq.~(\ref{eq:ddchi}) that $\chi_\ell$ is identically zero). Next, we rule out the exceptional case in which $\omega = 0$ and there existed a point $r_0\geq 0$ where $(R^2)'(r_0) = \chi_\ell'(r_0) = 0$. If this case occurred, successive differentiation of Eq.~(\ref{eq:ddR2_comb}) would yield
\begin{equation}
(R^2)'''(r_0) = 0,\qquad
\frac{(R^2)''''(r_0)}{R^2(r_0)} = \frac{2\ell+1}{2\pi}[\chi_\ell''(r_0)]^2.
\end{equation}
Further, evaluating Eq.~(\ref{eq:cons}) at $r = r_0$ one would obtain
\begin{equation}
-\frac{1}{R^2(r_0)} = \frac{2\ell+1}{8\pi}\left( \mu^2 - \frac{2\ell+1}{4\pi}\Lambda\chi_\ell(r_0)^2
 + \frac{\ell(\ell+1)}{R^2(r_0)} \right)\chi_\ell(r_0)^2,
\end{equation}
implying that $\chi_\ell(r_0)\neq 0$ and that the expression inside the parenthesis on the right-hand side must be negative. Eq.~(\ref{eq:ddchi}) would then imply that
\begin{equation}
a(r_0)\frac{\chi_\ell''(r_0)}{\chi_\ell(r_0)} < -\frac{2\ell+1}{4\pi}\Lambda\chi_\ell(r_0)^2 < 0,
\end{equation}
and hence $\chi_\ell''(r_0)\neq 0$ and $(R^2)''''(r_0) > 0$. It follows that any critical point of $R^2$ must be a strict minimum of $R^2$. However, since $R^2$ is convex there can be only one such critical point which is the one at the throat. Therefore, it follows from Eq.~(\ref{eq:ddR2_comb}) that $(R^2)''(r) > 0$ for all $r > 0$ and the theorem is proven.
\qed

\subsection{Numerical shooting algorithm}

Next, we describe a shooting algorithm which allows us to find asymptotically flat wormhole solutions from a given set of initial conditions at the throat by numerically integrating the equations outwards. As discussed above, there is only one free parameter to start the shooting procedure. Such parameter is the value of the scalar field at the throat, $\chi_\ell(0)$. 

We are looking for the desired solutions in the same spirit as the boson stars 
(see for instance \cite{Liebling:2012fv}), in which the solutions are 
parametrized by the value of the scalar field at the center of the configuration 
so that for each solution a set of discrete values for the frequency is found to 
satisfy the asymptotic flatness conditions, each with different number of nodes 
for the scalar field profile. Qualitatively, the same happens with 
the $\ell$-wormhole solutions discussed here. All the solutions reported in this 
article are those corresponding to the ground state, in which the scalar field 
$\chi_\ell$ has no nodes.

So for given values of $a(0)$, $\Lambda$, $\ell$, $\omega$, only one particular value of 
$\chi_\ell(0)$ picks the $\chi_\ell\rightarrow 0$ solution at infinity. We can 
see this in the approximation of the Klein-Gordon equation for large $r$. 
If we assume that $a(r)$ tends to unity and $R(r)$ to the coordinate $r$ fast enough, then 
Eq.~(\ref{assymp}) is satisfied, which is consistent with exponential decay of $\chi_\ell$ for large $r$ as long as $\mu^2 - \omega^2 > 0$. In a similar way we see that if $\chi_\ell$ is exponentially decaying at both infinities then, from~(\ref{eq:dda}) we obtain $\frac{d^2 a}{dr^2}+\frac{2}{r}\frac{d a}{dr}\approx 0$, which has solutions:
\begin{equation}\label{eq:ainf}
 a\approx B+\frac{A}{r},
\end{equation}
where $A$ and $B$ are constants. This is a particular simplification over Eq.~(\ref{eq:simeq}) that is useful in the numerical procedure. In particular $B$ will enter as a 
normalization factor, since we will ask for $B=1$, as required by the asymptotic condition~(\ref{eq:a_asym}).

As mentioned previously, $\chi_\ell(0)$ is used as the shooting parameter so the requirements needed to find a solution are those described in the previous paragraphs. Using the LSODA FORTRAN solver for  initial value problems of ordinary differential equations, we perform the integration of the system~(\ref{eq:ddchi}--\ref{eq:ddf}) starting at the value $r=0$ using steps of $\Delta r=1\times 10^{-6}$ until a final value is reached.
This finite value of the asymptotic boundary needs to be sufficiently large for the functions to reach 
their asymptotic behavior. Once the desired behavior of $\chi_\ell$ is obtained up to a precision of order $\Delta r$, the asymptotic values of $R$ and $a$ are adjusted by means of the transformation~(\ref{transf}) which leaves the system of equations invariant, where the parameter $B$ is chosen equal to the corresponding coefficient in Eq.~(\ref{eq:ainf}).

Examples are shown in Tables \ref{tab:solutions_lambda} and \ref{tab:solutions_omega}. Their physical implications are shown in section \ref{sec:properties}.
The 0-wormhole recovers the wormhole studied by Dzhunushaliev 
\textit{et al}. in~\cite{Dzhunushaliev:2017syc} for complex, massive and self-interacting ghost scalar fields. Our results match those of them as can be seen in the $\Lambda=4.0$ row in Table \ref{tab:solutions_lambda} when the following change of variables is performed:

\begin{equation}
r\mapsto\int_0^r\frac{dr}{\sqrt{a(r)}},\ \Lambda\mapsto\frac{\Lambda}{4},\ \chi_\ell\mapsto\sqrt{\frac{8\pi}{2\ell+1}}\chi_\ell,
\end{equation} 
which takes into account the differences in the definitions, nondimensionalization and the coordinate election. These authors also studied the case for a real scalar field in a previous work~\cite{Dzhunushaliev:2008bq}, which in fact corresponds to the $\omega=0$ results in this paper.

\begin{table}
	\begin{ruledtabular}
		\begin{tabular}{l lll | lll | lll}  
			\toprule
			$\omega=0$&\multicolumn{3}{c}{$\ell=0$}&\multicolumn{3}{c}{$\ell=1$}&\multicolumn{3}{c}{$\ell=2$} \\
			\cmidrule(r){2-10}
			&$\chi_\ell(0)$    & $R(0)$ & $a(0)$&$\chi_\ell(0)$    & $R(0)$ & $a(0)$&$\chi_\ell(0)$    & $R(0)$ & $a(0)$ \\
			\colrule
			$\Lambda=0.5$&6.26 & 1.07 & 1.40 & 3.94 & 1.72 & 7.43 & 3.08 & 2.71 & 20.21 \\
			$\Lambda=0.7$&4.91 & 1.75 & 0.80 & 3.21 & 1.97 & 2.59 & 2.58 & 2.81 & 7.04 \\
			$\Lambda=1.0$&3.90 & 2.79 & 0.58 & 2.48 & 2.70 & 1.03 & 2.08 & 3.14 & 2.51 \\
			$\Lambda=4.0$&1.81 & 13.21 & 0.41 & 1.05 & 13.14 & 0.42 & 0.82 & 13.00 & 0.44 \\
		\end{tabular}
	\end{ruledtabular}
	\caption {Central values of the field $\chi_\ell$ and metric functions $R$ and $a$ for several values of $\Lambda=0.5,0.7,1.0,4.0$ and $\ell=0,1,2$ with $\omega=0$.}\label{tab:solutions_lambda}
\end{table}

\begin{table}
	\begin{ruledtabular}
		\begin{tabular}{l lll |l lll |l lll}  
			\toprule
			$\Lambda=0.7$&\multicolumn{3}{c}{$\ell=0$}& &\multicolumn{3}{c}{$\ell=1$}& &\multicolumn{3}{c}{$\ell=2$} \\
			\cmidrule(r){2-12}
			&$\chi_\ell(0)$    & $R(0)$ & $a(0)$&&$\chi_\ell(0)$    & $R(0)$ & $a(0)$&&$\chi_\ell(0)$    & $R(0)$ & $a(0)$ \\
			\colrule
			$\omega=0.1$&4.88 & 1.77 & 0.83 & $\omega=0.3$ & 3.18 & 2.01 & 2.60 & $\omega=0.8$ & 2.56 & 2.85 & 7.03 \\
			$\omega=0.2$&4.64 & 2.01 & 1.07 & $\omega=0.5$ & 2.95 & 2.34 & 2.82 & $\omega=1.0$ & 2.36 & 3.28 & 7.13 \\
			$\omega=0.3$&4.06 & 3.01 & 2.25 & $\omega=0.8$ & 2.45 & 3.67 & 4.49 & $\omega=1.6$ & 1.94 & 5.01 & 9.77 \\
		\end{tabular}
	\end{ruledtabular}
	\caption {Central values of the field $\chi_\ell$ and metric functions $R$ and $a$ for several values of $\omega$ and $\ell=0,1,2$ with $\Lambda=0.7$.}\label{tab:solutions_omega}
\end{table}

\subsection{Energy density, mass and curvature scalars}

In order to help interpreting the solutions presented in the next section, 
we discuss several scalar quantities, like the energy density $\rho$ of the 
ghost field measured by static observers, the Misner-Sharp mass function 
and the scalars related with the curvature of the spacetime, such as the Ricci 
scalar $R_s$ and the Kretschmann scalar $K$. These quantities will turn out to 
be helpful for understanding the features of the ghost field and its action on 
the geometry.

Explicitly, the function $\rho$, associated with the density of the ghost field, is given by
\begin{equation}
\rho=-\frac{\tensor{T}{^t_t}}{c^2}
=\left(\frac{c^2}{8\pi\,G}\right)\,\hat{\rho},
\end{equation}
where $\hat{\rho}$ is defined by
\begin{equation}
\hat{\rho}=-(2\ell+1)\left[a\,\chi_\ell'^2+\left
(\frac{\ell(\ell+1)}{R^2}+\mu^2-\frac{2\ell+1}{4\pi}\Lambda\chi_\ell^2+\frac{
\omega^2}{a}\right)\chi_\ell^2\right].
\label{eq:rhohat}
\end{equation}
A striking feature of the wormhole solutions is that despite the presence of the exotic matter which violates the null energy condition everywhere, the density may still be \emph{positive} at the throat,\footnote{Note that the violation of the null energy condition implies the violation of the weak energy condition, which means that there exists at least one observer which measures negative energy density. Our example shows that this observer does not necessarily need to be a static one.}
as will be shown in the numerical examples discussed  in the next section. In fact, using Eq.~(\ref{eq:cons}) one can obtain the following simple expression for $\hat{\rho}$ at the throat:
\begin{equation}
\hat{\rho}(0) = \frac{8\pi}{b^2} - 2(2\ell+1)\frac{\omega^2}{a(0)}\chi_\ell(0)^2,
\end{equation}
which shows explicitly that for those solutions with $\omega = 0$ the energy density is indeed positive near the throat. The plots in the next section show that this behavior also holds for other solutions with small enough values of $\omega^2$.

The total (ADM) mass of the wormhole configurations can be computed from the 
asymptotic limit $M_\infty = \lim\limits_{r\to \infty} M(r)$ of the Misner-Sharp 
mass function~\cite{cMdS64} $M(r)$, defined by
\begin{equation}
\frac{2GM}{c^2} = R\left[ 1 - g^{\mu\nu}(\nabla_\mu R)(\nabla_\nu R) \right]
 = R\left( 1 - a R'^2 \right).
\end{equation}
From Eqs.~(\ref{eq:simeq},~\ref{eq:ainf}) one obtains $GM_\infty/c^2=c_0/2=-A/2$. Alternatively, using Eqs.~(\ref{eq:ddf},~\ref{eq:cons}) one also obtains $M' = 4\pi \rho R^2 R'$ which can be integrated to
\begin{equation}
M(r) = \frac{c^2}{2G}\left( b 
 + \int\limits_0^r \hat{\rho}(\bar{r}) R^2(\bar{r}) R'(\bar{r}) d\bar{r} \right),
\end{equation}
with $b = R(0)$ the throat's areal radius. As long as $\hat{\rho}$ is positive near 
the throat, the mass function increases as one moves away from the throat. 
However, $M$ decreases as soon as $\hat{\rho}$ becomes negative, so that 
solutions which have either sign of the total mass are possible.
This is shown in Table~\ref{tab:mass}, where values of the total mass $M_\infty$
for our wormhole were computed taking several values of $\ell$ and fixing the values of all other parameters.

\begin{table}
		\begin{tabular}{c|c}  
			\toprule
			$\omega=0$, $\Lambda=1$ & $M_\infty\ (m_{\text{pl}}^2/m_\Phi)$\\
			\colrule
			$\ell=0$ & $0.677$\\
			$\ell=1$ & $0.224$\\
			$\ell=2$ & $-1.25$\\
			$\ell=3$ & $-4.69$\\
		\end{tabular}
	\caption {Total mass values for $\ell$-wormholes with $\Lambda=1$, $\omega=0$ and $\ell=0,1,2,3$.}\label{tab:mass}
\end{table}

The Ricci scalar, $R_s={R^\mu}_{\mu}$, associated with the geometry given by 
Eq.~(\ref{eq:metric}) has the form 
\begin{equation}
 R_s = - a'' - 2\,a\,\frac{{R^2}''}{R^2} + a\,\left(\frac{{{R^2}'}}{R^2}\right)^2 + \frac{2}{R^2}\,\left(1 - a'\,{R^2}'\right).
 \end{equation}
A further commonly used curvature measure is the Kretschmann scalar, defined by 
$K=R^{\mu\nu\sigma\tau}\,R_{\mu\nu\sigma\tau}$. For the metric under 
consideration, Eq.~(\ref{eq:metric}), the Kretschmann scalar has the following 
explicit form:
\begin{equation}
K=a'' + 2\,\left(\frac{a\,{R^2}''}{R^2}\right)^2 + 2\,a\,\frac{{R^2}''\,{R^2}'}{R^4}\,\left(a' - \frac{a\,{R^2}'}{R^2}\right) + \frac{4}{R^4}
+ \left[\frac{3}{4}\,\left(\frac{a\,{R^2}'}{R^2}\right)^2 - \frac{{a^2}'\,{R^2}'}{2\,R^2} + {a'}^2 - 2\,\frac{a}{R^2}\right]\,
 \left(\frac{{R^2}'}{R^2}\right)^2. 
\end{equation}

All these quantities will turn out to be helpful when understanding the role played by the several parameters of the solution in the geometry and in the dynamics of the bodies moving on it.

From Einstein's equations, Eq.~(\ref{eqs:Eins}), we have that 
$R_s=-\frac{8\,\pi\,G}{c^4}\,T$ with $T$ the trace of the stress-energy-momentum tensor. Numerical experiments show that the behavior of the stress-energy-momentum tensor components in the throat region are similar to each other, and thus $R_s \approx \frac{8\,\pi\,G}{c^2}\,\rho$, as seen in the actual solutions. That is, the Ricci scalar goes as the 
density, irrespective of its character, exotic or usual matter. We will discuss
this fact in more detail in the explicit cases that we present below.

\section{Numerical wormhole solutions}
\label{sec:properties}

Following the procedure described above, we are able to obtain several solutions to the Einstein-Klein-Gordon system, given the
four parameters, namely $\mu, \omega, \Lambda$, and $\ell$. We will present the solutions 
first for trivial values of the angular momentum parameter, $\ell=0$,
and vary the self-interaction parameter $\Lambda$, while 
keeping the oscillation frequency $\omega$ fixed and then we explore the properties of the 
solution for some values of $\omega$ maintaining $\Lambda$ fixed, as was done in ~\cite{Dzhunushaliev:2017syc, Dzhunushaliev:2008bq}.
Next, we repeat the study for different values of $\ell$. In all our solutions 
presented in this work, we keep the mass of the scalar field $\mu$ fixed.
These experiment allow us to have a better understanding on the role that each parameter plays in determining the geometry of the solutions. 

All the solutions presented are asymptotically flat, and are generated by 
looking for a  solution of the ghost scalar field, once the parameters 
$\ell, \Lambda$ and $\omega$ are chosen. We fix the value of  mass parameter 
$\mu$ to one, and the distance scale of the solution is given by the 
dimensionless parameter $\hat{r}=\mu r$.
Also, from Eq.~(\ref{eq:throat_cons}) we see that the size of the wormhole throat $R(0)$ is
given by
\begin{equation}
R(0)=\left(\frac{\ell(\ell+1)+\frac{8\pi}{(2\ell+1)\chi_\ell^2(0)}}{\frac{2\ell+1}{4\pi}\Lambda\chi_\ell^2(0)+\frac{\omega^2}{a(0)}-\mu^2}\right)^{1/2}. \label{eq:throat}
\end{equation}

\begin{figure}[!ht]
\includegraphics[width=0.45\columnwidth]{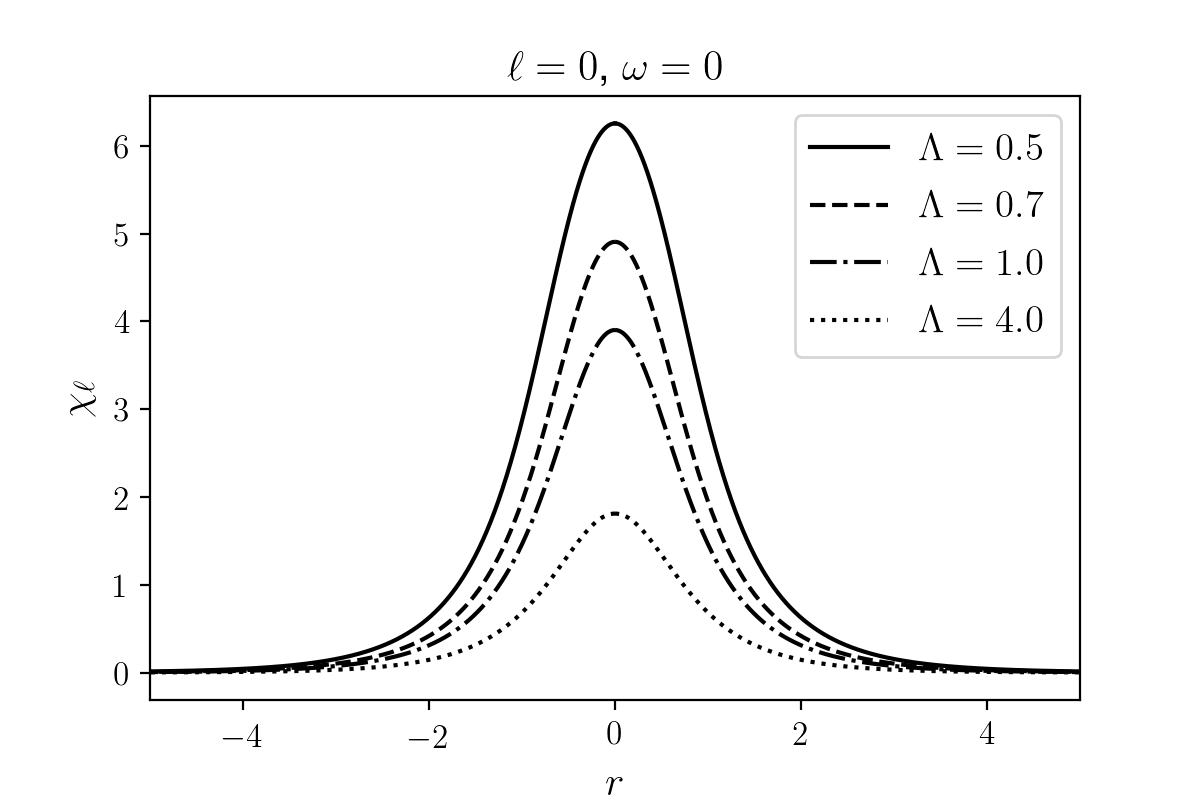}\\
\caption{Solitonic profile for the 0-wormhole, for $\Lambda\,\in\,[0.5, 4.0]$, and $\omega=0$.}
 \label{fig:Chi}
\end{figure}

In Fig.~\ref{fig:Chi}, we present this localized solution, for the case $\ell=0$, $\omega=0$, for several values of $\Lambda$. All the other solutions with $\ell>0$ are localized as well. Notice how the amplitude of the pulse decreases as the value of the
self-interaction parameter $\Lambda$ increases.

In our experiments, we see that the ghost density, in order to form a 
wormhole, is distributed in such a way that it has a positive value in the 
region of the throat, and then it starts to have larger concentrations of negative ghost 
density on both sides of the throat, as shown in Fig.~\ref{fig:rho_0_ch}. From the geometric perspective, as suggested above, the profile of the Ricci scalar follows the density one and has a convex region at the throat, surrounded by concave zones, see Fig.~\ref{fig:R01_K01}. 

Also we will show that, in general, as can be seen in 
Fig.~\ref{fig:rho_0_ch}, the action of the self-interaction parameter, 
$\Lambda$, smooths out this behavior of the exotic density and spacetime 
interaction. Indeed, the scalar field, at least the massive ghost field, 
possess a radial pressure that creates the throat and then the spacetime 
strongly reacts generating regions of negative density; it is the role of 
the self-interaction term to smooth down such reaction and allows to keep the 
wormhole throat open with smaller amount of ghost density. Conversely, as the self-interaction parameter $\Lambda$ becomes smaller, the metric coefficient $a$, the curvature scalars and density at the throat become more and more localized, an observation which is compatible with the result in Theorem~1 where we have shown that the solutions cease to exist for $\Lambda = 0$.

\subsection{0-wormhole}
\label{sec:0lw}

\begin{figure}[!ht]
\includegraphics[width=0.45\columnwidth]{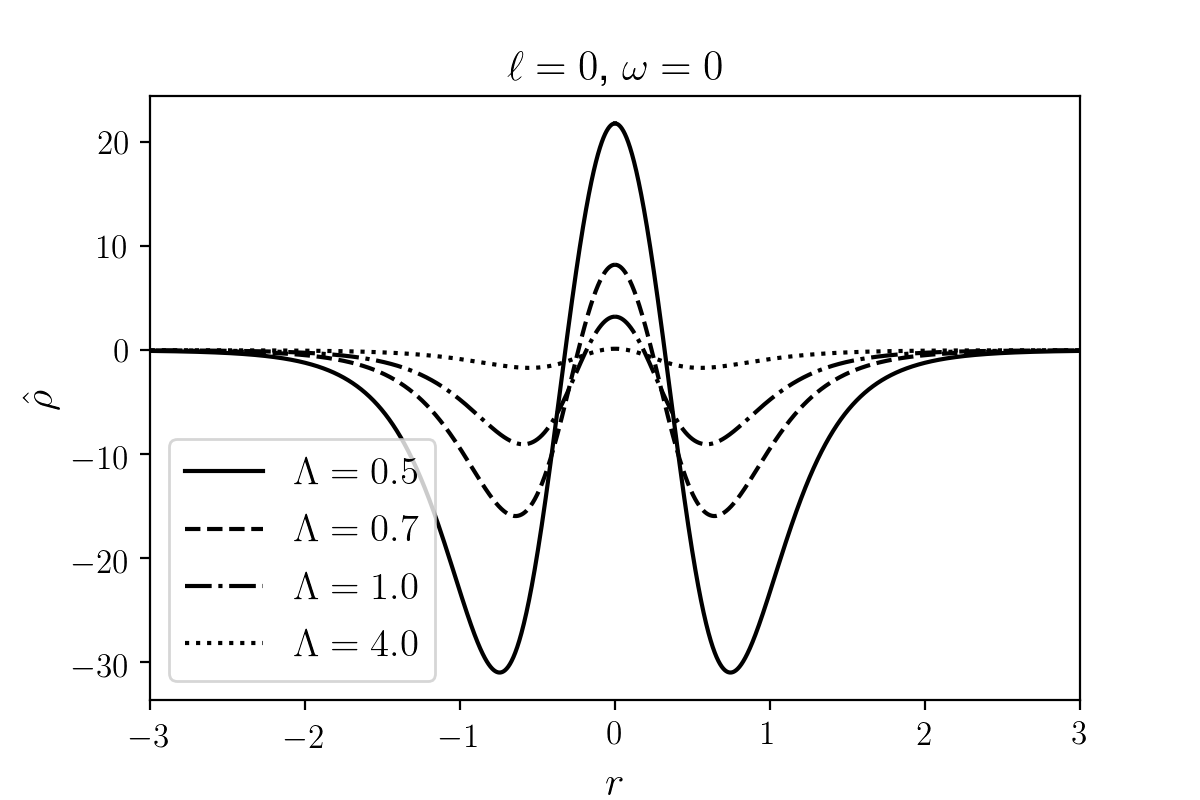}\\
\caption{Density profile for the $0$-wormhole, for $\Lambda\,\in\,[0.5, 4.0]$ 
and $\omega=0$.}
 \label{fig:rho_0_ch}
\end{figure}

We start our discussion for the case with vanishing angular momentum, {\it i. e.} $\ell=0$. Setting also $\omega$ equal to zero for the moment, we start by sweeping a range of values for the self-interaction parameter, $\Lambda$. The corresponding results for the scalar field and the density profile are shown in Figs.~\ref{fig:Chi} and~\ref{fig:rho_0_ch}, respectively.
As mentioned above, the ghost density has regions of positive magnitude near the throat, and 
regions with negative density which tend to zero from below in the asymptotic region. 

\begin{figure}[!ht]
\includegraphics[width=0.45\columnwidth]{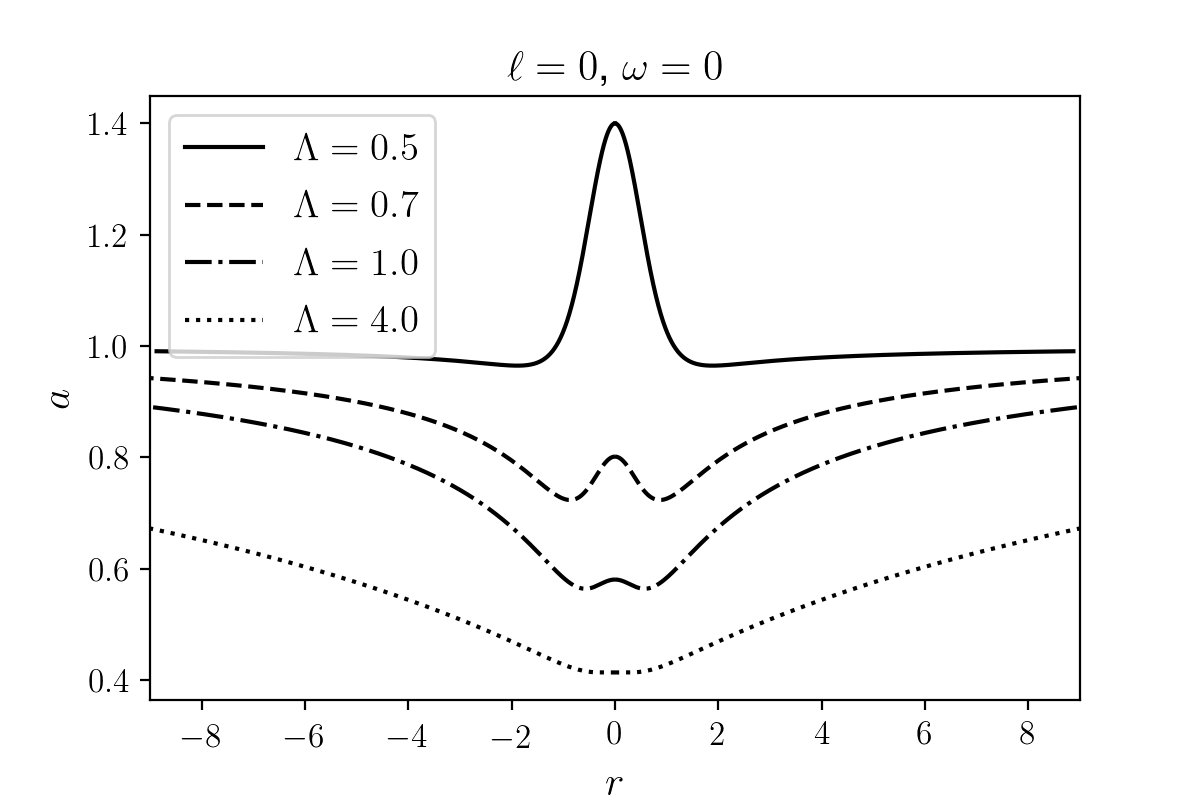}\hspace{0.5cm}
\includegraphics[width=0.45\columnwidth]{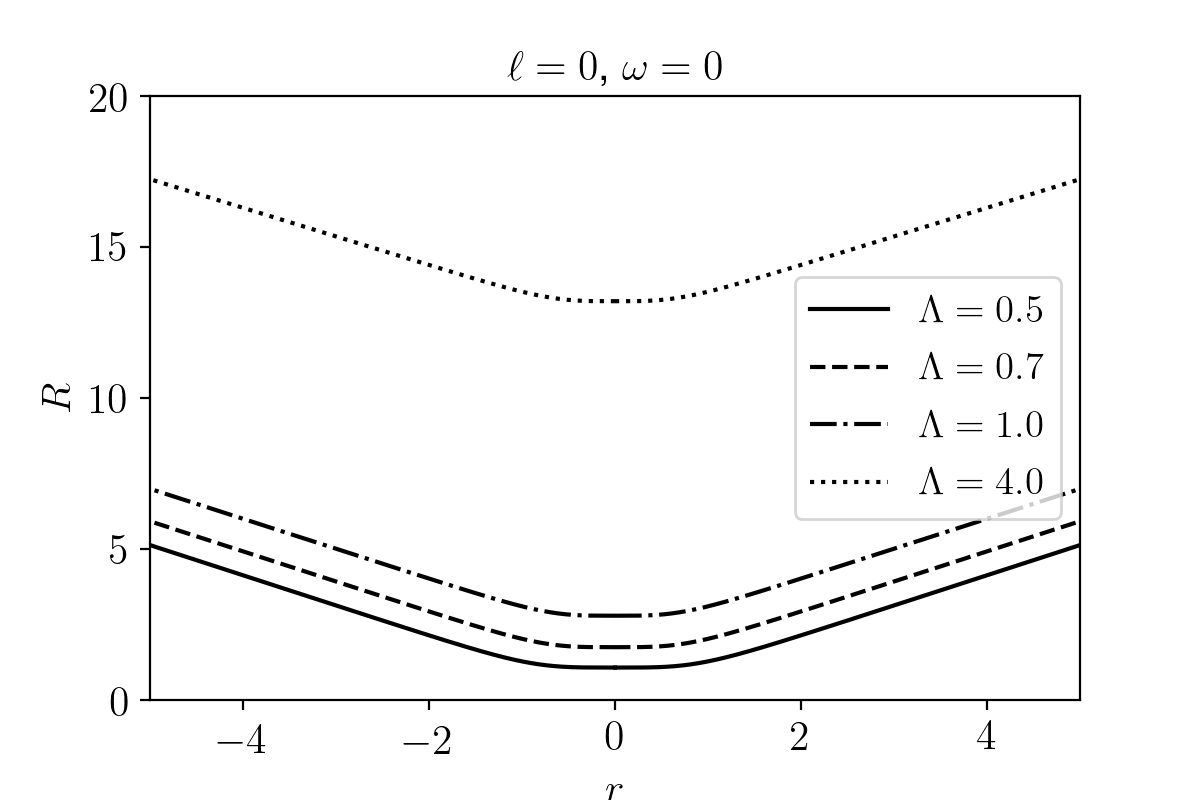}
\caption{Metric coefficients for $\Lambda\,\in\,[0.5, 4.0]$ and $\omega=0$.}
 \label{fig:a01_R201}
\end{figure}

The corresponding metric coefficients, $a(r)$ and $R(r)$ are shown in Fig.~\ref{fig:a01_R201}. Notice how the metric coefficient $a(r)$ shows concave regions which will determine a similar behavior in the effective potential of the spacetime, which in turn will imply the existence of
particles moving on bound trajectories. Again, the effect of the 
self-interaction parameter is to smooth  out the concavity of the metric functions.

\begin{figure}[!ht]
\includegraphics[width=0.45\columnwidth]{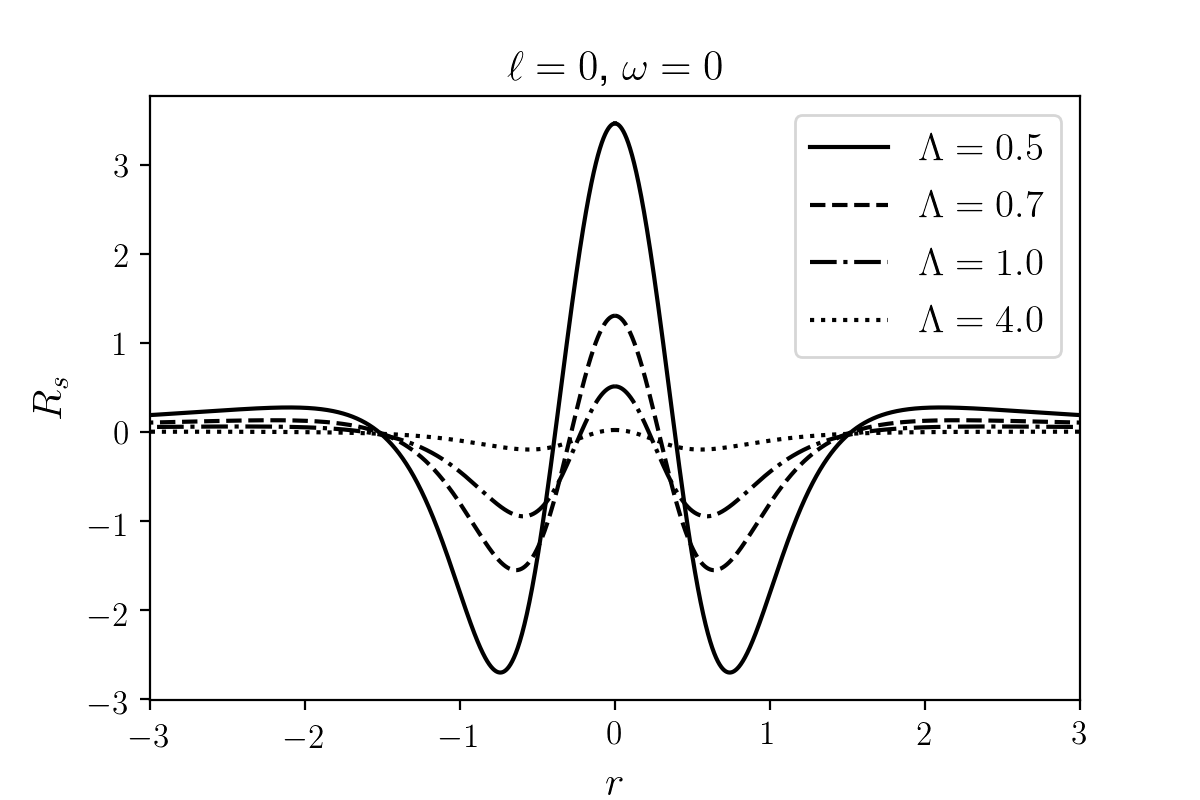}\hspace{0.5cm}
\includegraphics[width=0.45\columnwidth]{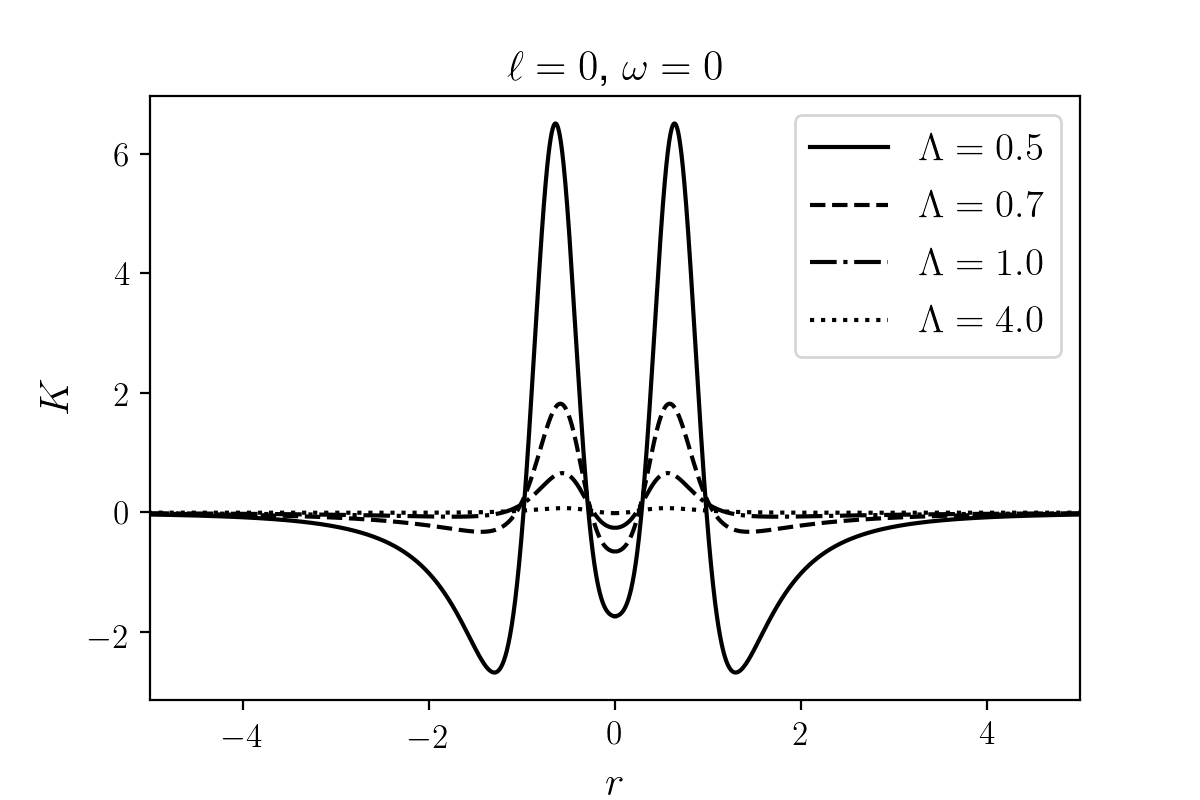}
\caption{Ricci and Kretschmann scalars for $\Lambda\,\in\,[0.5, 4.0]$ and $\omega=0$.}
 \label{fig:R01_K01}
\end{figure}

Regarding the curvature scalars, as expected, the Ricci scalar $R_s$ has a 
behavior which follows the one of the density, with regions of positive values 
and then valleys with negative values of the curvature as we can see in Fig.~\ref{fig:R01_K01}. The Kretschmann scalar, 
however, is very different and shows two peaks of positive values and they 
decrease as the self-interaction parameter grows, 
and the central one is negative in the region of the throat, surrounded 
by bumps.

\begin{figure}[ht]
	\includegraphics[width=0.45\columnwidth]{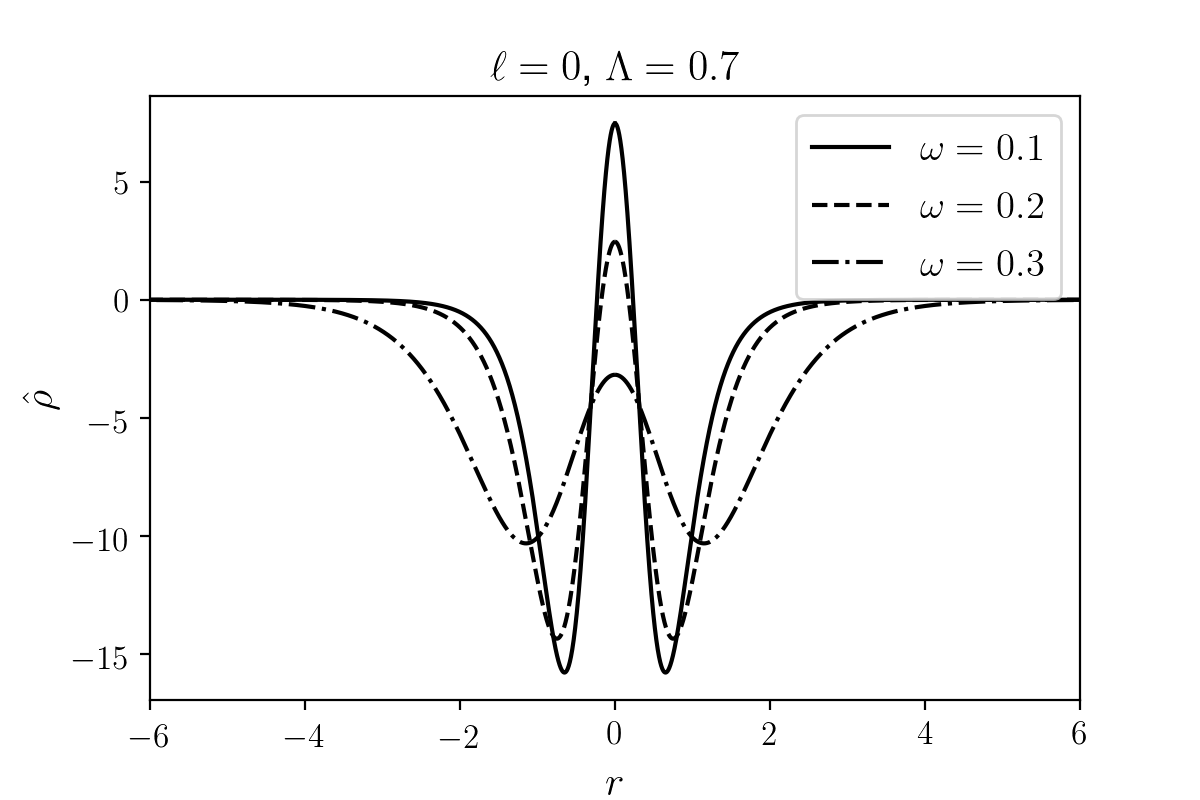}\hspace{0.5cm}
	\includegraphics[width=0.45\columnwidth]{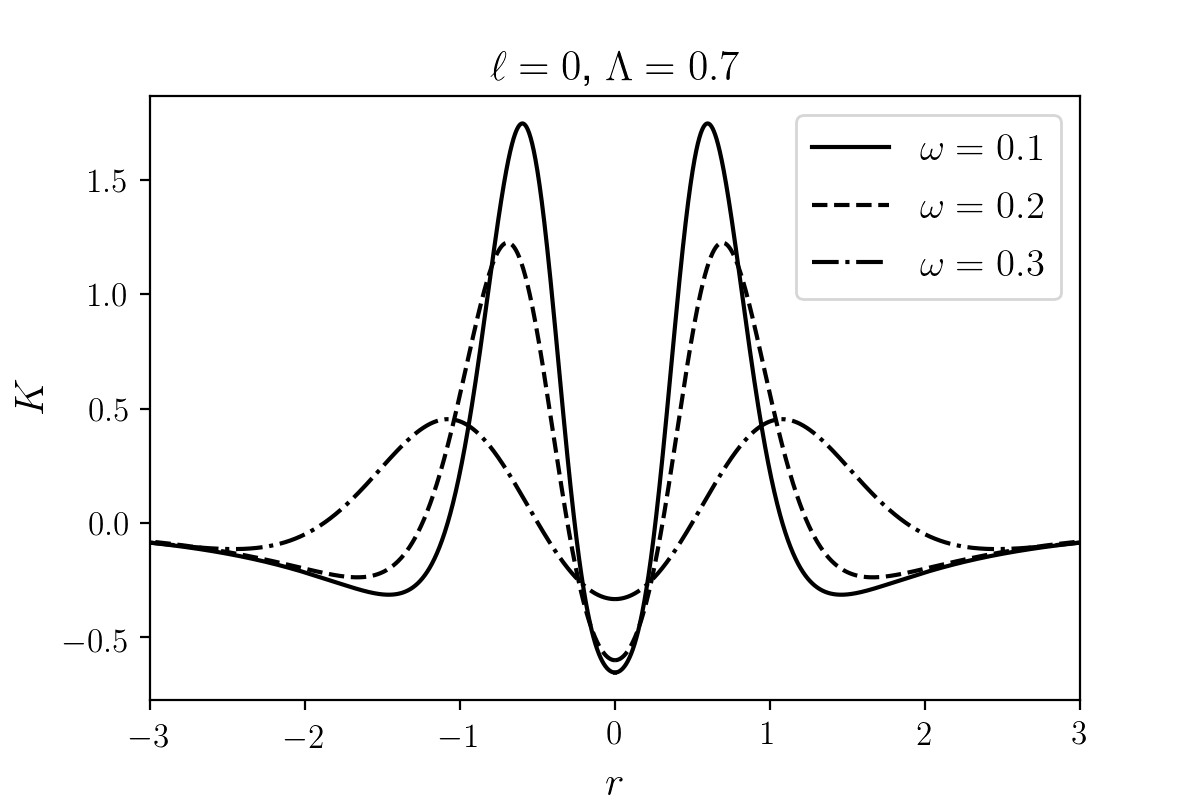}
	\includegraphics[width=0.45\columnwidth]{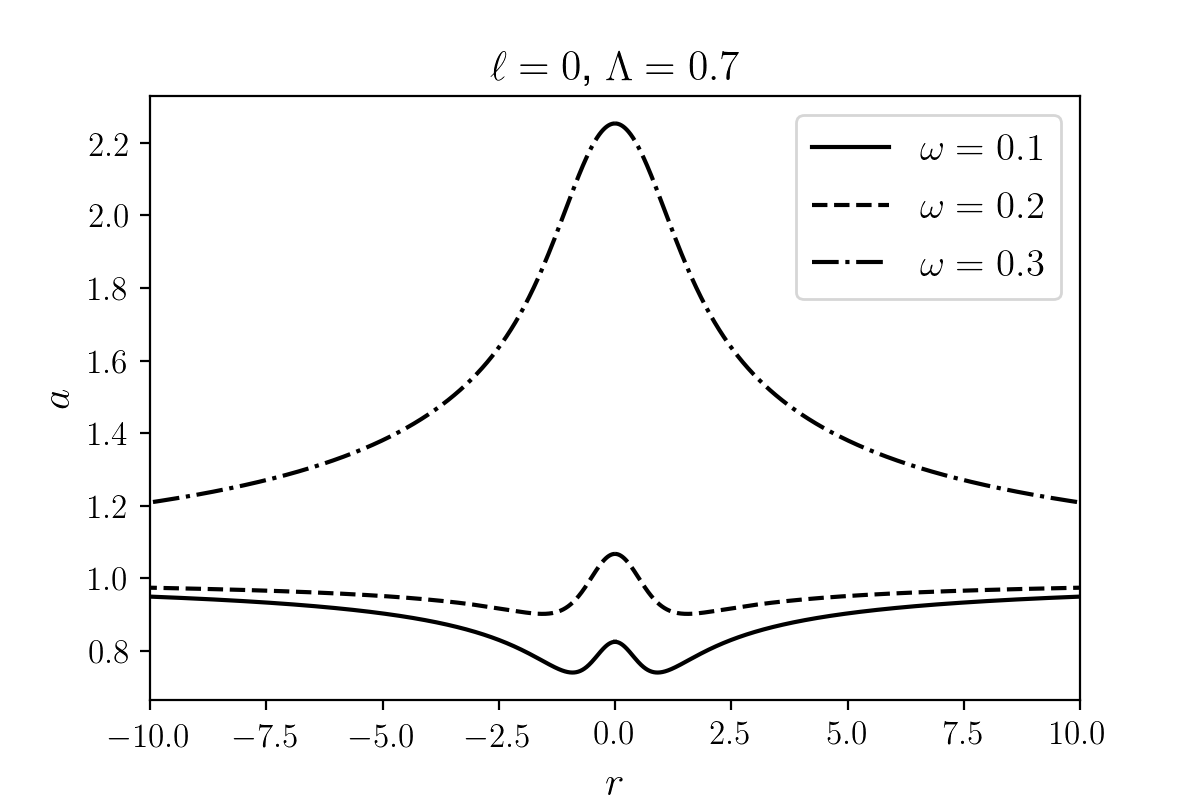}
	\caption{Density profile, curvature scalar and metric function $a$ for the 0-wormhole, $\Lambda=0.7$ and $\omega=0.1, 0.2, 0.3$.}
	\label{fig:wrho01_wK01}
\end{figure}

The next step is to increase the parameter $\omega$ keeping $\ell=0$ and the self-interaction parameter $\Lambda=0.7$ fixed. We show in  Fig.~\ref{fig:wrho01_wK01} the corresponding density and Kretschmann scalar for three non-zero values, $\omega=0.1, 0.3, 0.5$, of the frequency. Notice the difference between the behavior of the Kretschmann scalar, in which a larger value of $\omega$ gives the effect of increasing 
the central value, acting in the same way as the parameter $\Lambda$ discussed 
above.

\pagebreak

\subsection{$\ell$-wormhole}
\label{sec:lw}

\begin{figure}[!ht]
	\includegraphics[width=0.45\columnwidth]{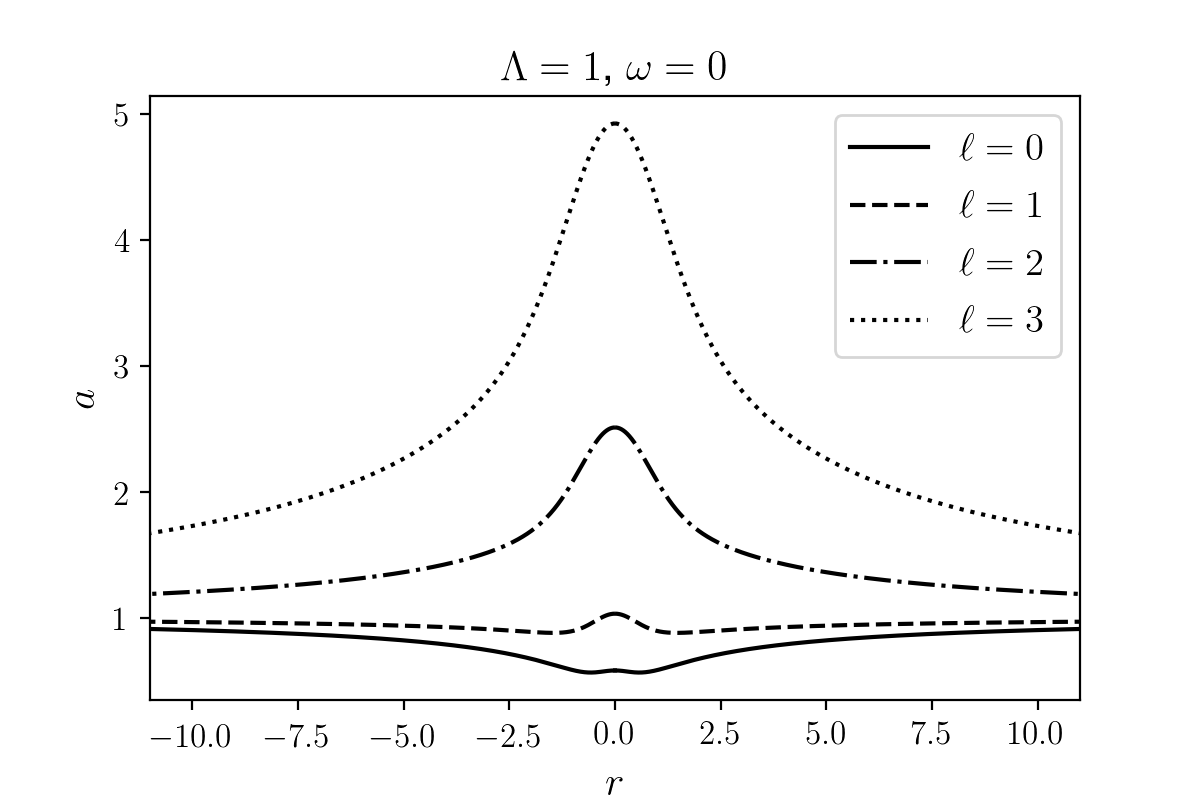}
	\includegraphics[width=0.45\columnwidth]{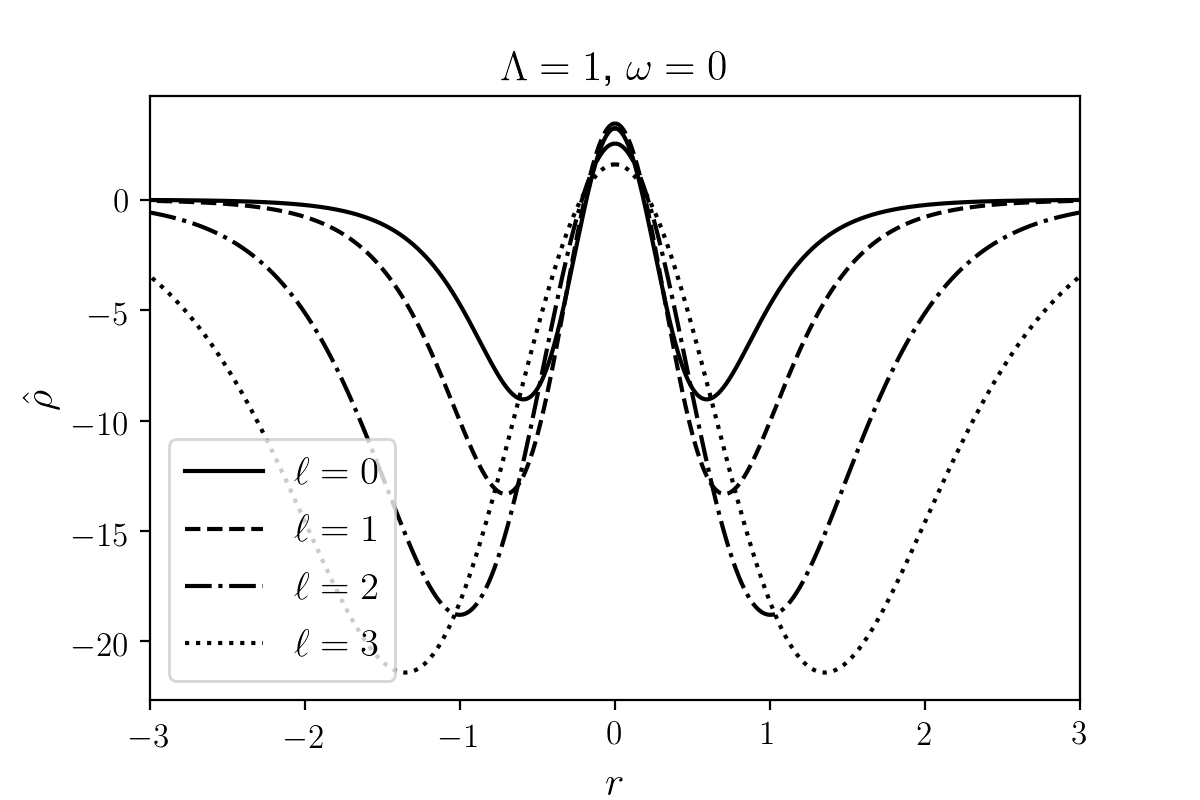}
	\includegraphics[width=0.45\columnwidth]{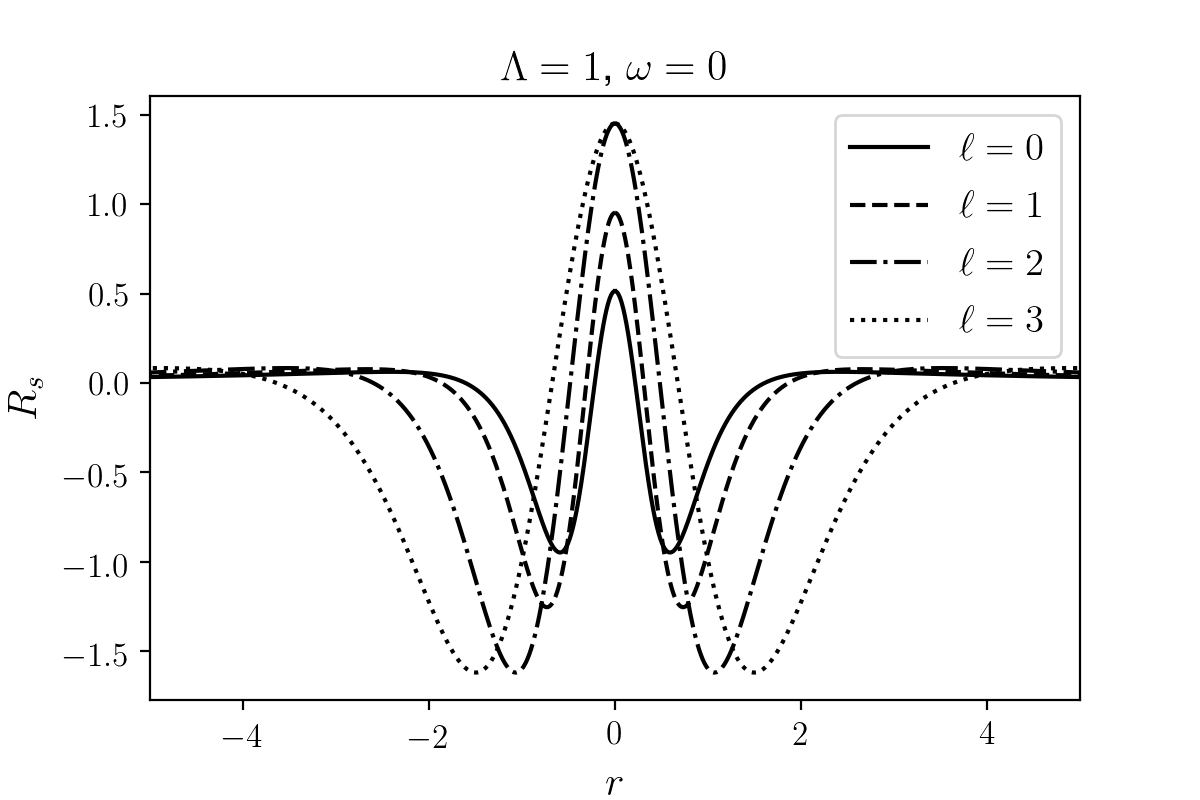} \includegraphics[width=0.45\columnwidth]{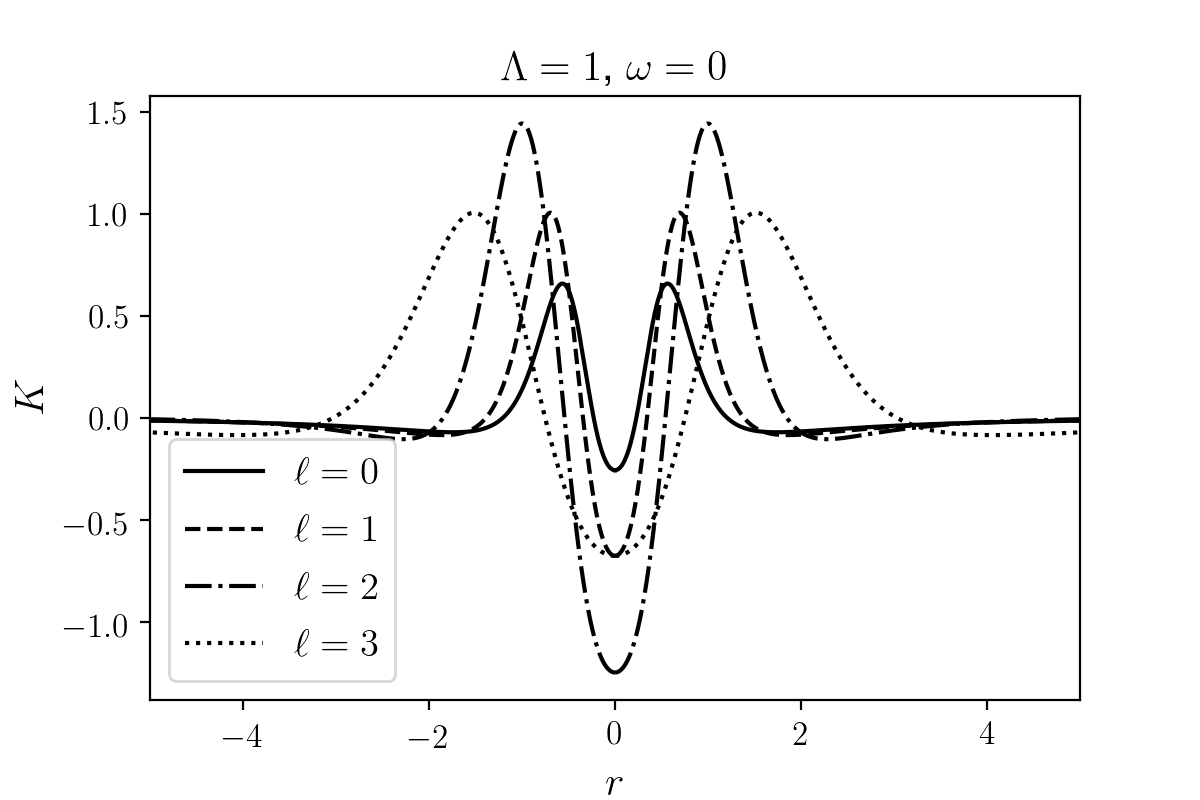}
	\caption{Metric function $a(r)$, density profile $\hat{\rho}$ and $R_s$, $K$ scalars for $\ell\geq0$, $\Lambda=1.0$ and $\omega=0$.}
	\label{fig:ellFigs}
\end{figure}

\begin{figure}[!ht]
	\includegraphics[width=0.6\columnwidth]{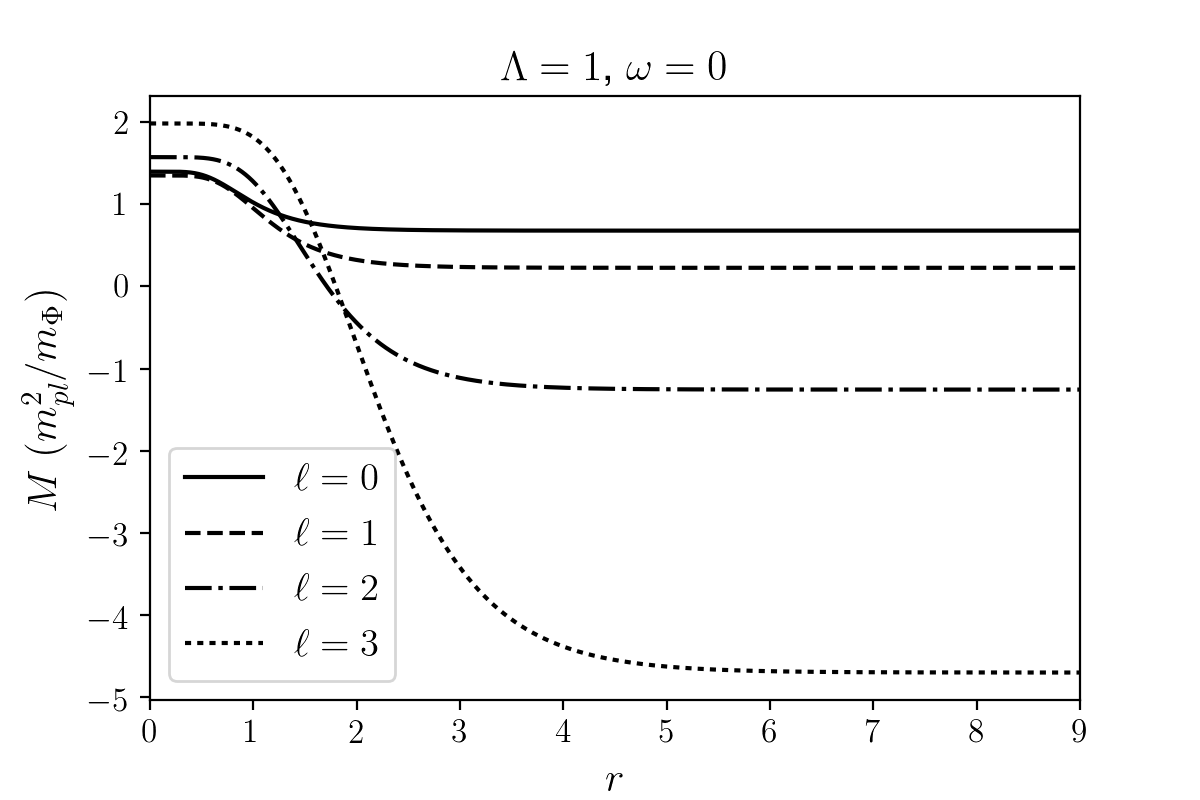}\\
	\caption{Mass function $M(r)$ for the parameters $\ell\geq 0,\;\Lambda=1\;\text{and}\;\omega=0$.}
	\label{fig:mass}
\end{figure}

In this section we present the behavior of the $\ell$ parameter and the effect 
on the metric functions and the curvature scalars. For the latest we can see in 
Fig.~\ref{fig:ellFigs} that an increment on the $\ell$ parameter increases the 
central peak for the Ricci scalar and decreases the central bump for the 
Kretschmann scalar. The case for $a(r)$ is quite different: while for $\ell=1$ 
the two minima are 
still present, for larger values of the $\ell$ parameter the central peak is 
increased and the minima disappear. 
The presence of a minimum (or two in 
this case) also corresponds to positive total masses as can be verified in 
Table~\ref{tab:mass} and Fig.~\ref{fig:mass}, consequently, its absence 
corresponds to negative masses. This is a general property of all solutions 
given the asymptotic behavior of $a$ (see Eq.~\ref{eq:ainf}).

\begin{figure}[!ht]
	\includegraphics[width=0.45\columnwidth]{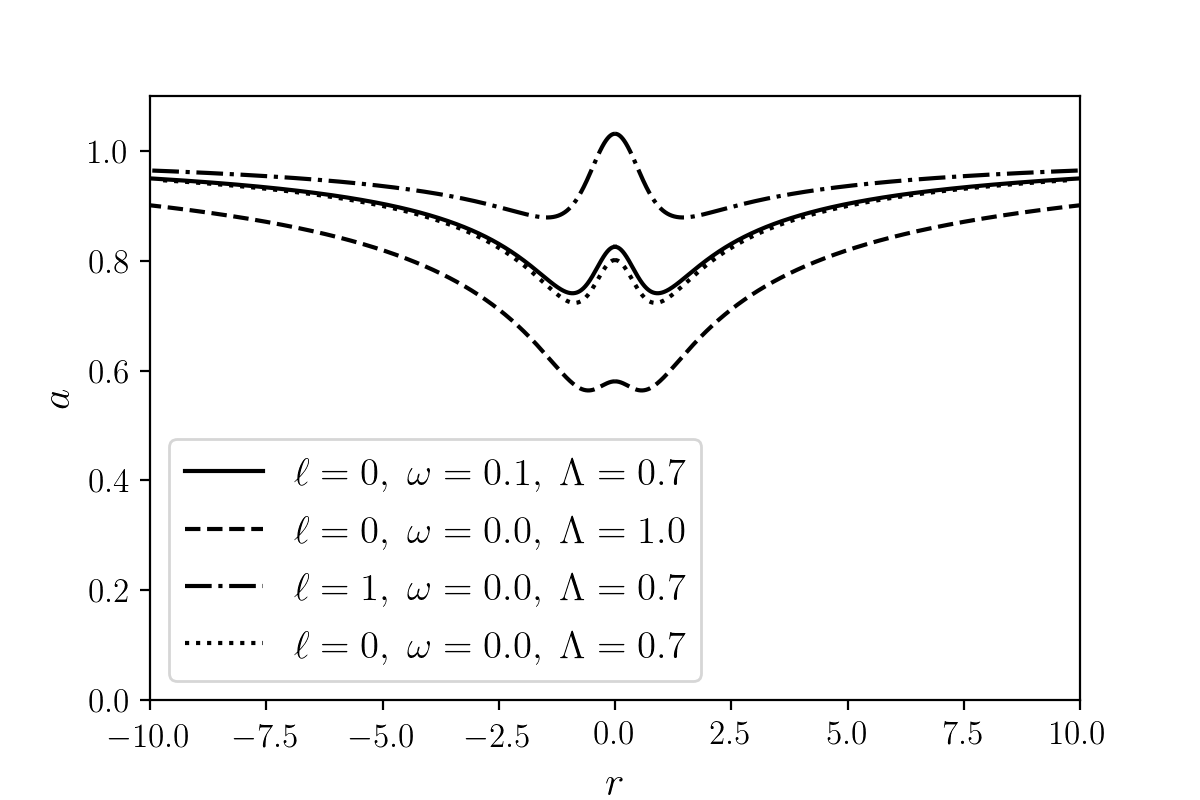}\\
	\caption{Metric function $a$ for different values of the parameters $\ell,\;\omega\;\text{and}\;\Lambda$.}
	\label{fig:a_lwh}
	\end{figure}

On the other hand, as is shown in Fig.~\ref{fig:a_lwh}, the increment of the $\omega$ parameter plays a role quite similar to the one made by the $\ell$ parameter: an increase on the former elevates the central peak on the metric function $a(r)$. 

Moreover, as can be seen from a comparison of Figs.~\ref{fig:a01_R201} and \ref{fig:ellFigs}, 
the effect of the $\Lambda$ parameter on the metric coefficient $a$ is opposite to the one generated by the $\ell$ parameter on that metric coefficient. Indeed, for small values of $\Lambda$, the metric coefficient $a$ has a global maximum at the throat, while for large values of this $\Lambda$, the metric coefficient $a$ only has a local maximum. Thus, for small values of $\Lambda$, the $\ell$ parameter is not able to change the qualitative behavior of the metric coefficient, while for larger values of $\Lambda$, the appearance of the local maximum is recovered or enhanced with the parameter $\ell$. This fact will have consequences on the effective potential and the geodesic motion of particles, as discussed below.

\section{Embedding diagrams and geodesic motion}
\label{sec:nembedding}

In order to gain a better understanding of the configurations described by the scalar field and the geometry in the vicinity of the throat, in this section we discuss the embedding procedure and geodesic motion. Because the metric~(\ref{eq:metric}) is static and spherically symmetric, it is sufficient to analyze the induced geometry on a $t=$ constant and $\theta = \pi/2$ slice, described by the two-metric
\begin{equation}\label{eq:wormslice}
d\Sigma^2 =  a^{-1}dr^2+R^2 d\varphi^2 \ .
\end{equation}
In order to visualize this geometry as a two-dimensional surface embedded in 
three-dimensional flat space we shall employ cylindrical coordinates ($\rho$, 
$\varphi$, $z$). The metric for a flat space in these coordinates is
\begin{equation}\label{eq:flatmetric}
dS^2=d\rho^2 +\rho^2d\varphi^2+dz^2 \ . 
\end{equation}
We seek for the functions $\rho(r)$ and $z(r)$, specifying a surface with the same 
geometry as the one described by the metric (\ref{eq:wormslice}).

The line element for the embedding surface will be
\begin{equation}
d\Sigma^2 = \left[\left( \frac{dz}{dr} \right)^2 + \left( \frac{d\rho}{dr} 
\right)^2   \right]dr^2+\rho^2d\varphi^2 \ ,
\end{equation}
if the following conditions are satisfied:
\begin{equation} \label{eq:embedd_aux1}
\rho = \,R \ ,
\end{equation}
and 
\begin{equation}\label{eq:embedd_aux2}
\left( \frac{dz}{dr} \right)^2 + \left( \frac{d\rho}{dr} 
\right)^2=\frac{1}{ a} \ .
\end{equation}

Using the expression~(\ref{eq:embedd_aux1}) to calculate $\frac{d\rho}{dr}$, 
Eq.~(\ref{eq:embedd_aux2}) gives the following differential equation for $z(r)$:
\begin{equation}
\frac{dz}{dr} = \left[\frac{1}{ a}- \frac{1}{4}\frac{({R^2}')^2}{R^2} 
\right]^{1/2} \ .
\end{equation}

Integrating this equation gives the function $z=z(r)$; in order to plot it in an Euclidean space, we need to find $r$ as a function of $\rho$. However, it is not possible to express this function $r = r(\rho)$ in closed form because $R$ was found numerically. Nevertheless, one can obtain $r = r(\rho)$ numerically from~(\ref{eq:embedd_aux1})  and finally get $z=z(\rho)$.

\begin{figure}[ht]
\includegraphics[width=0.5\columnwidth]{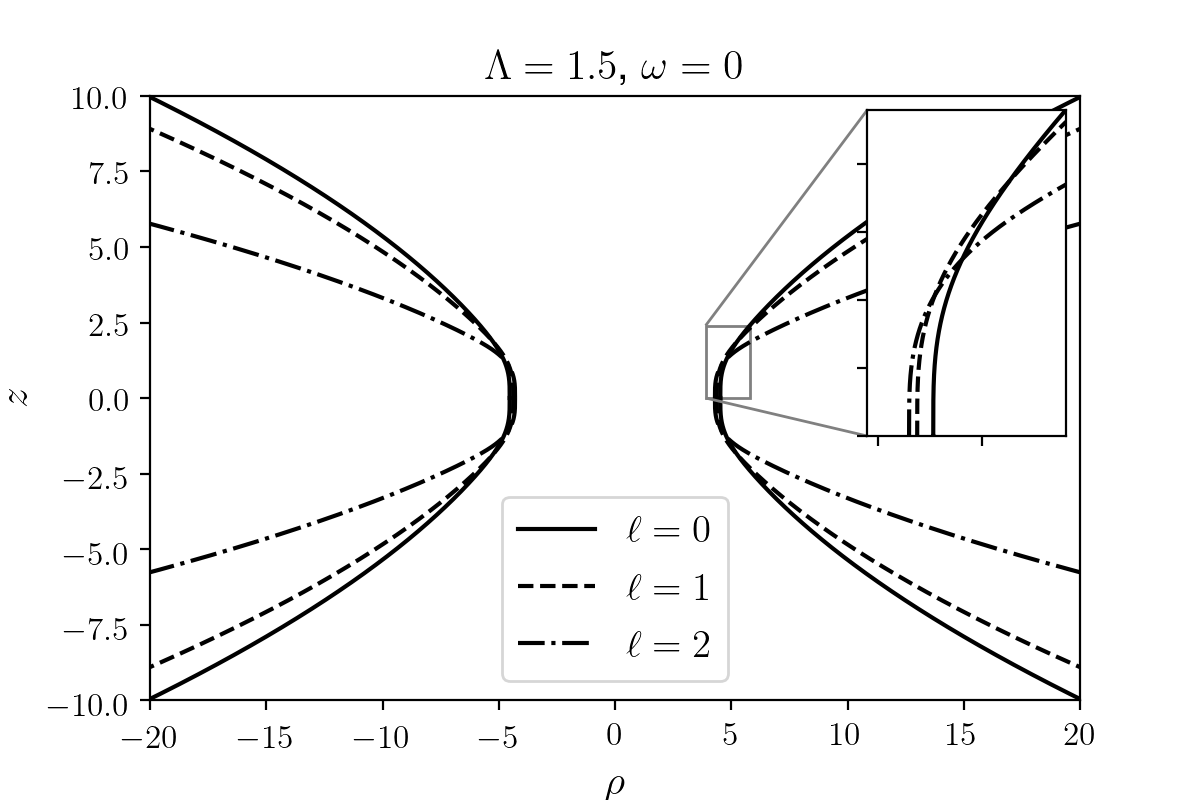}
 \caption{Embedding of the different $\ell$-wormholes, for $\Lambda=1.5, \omega=0$. The complete embedding diagram is obtained by rotating this figure about the $z$ axis. In the enlarged picture we underline the change of the value of the throat radius for the given values of $\ell$.
 } 
 \label{fig:embedding}
\end{figure}

In Fig.~\ref{fig:embedding} we show the visualization of this embedding. 
It is seen that as $\ell$ increases from 0 to 2, the profile of of the embedding representing the wormhole's geometry becomes more and more curved (which is analogous to the increase of $|R_s|$ and $|K|$ shown in Fig. 10) , with a slight decrease in the throat's radius.
%

\subsection{Geodesic motion}
\label{sec:geodesics}

In order to describe the motion of the particles in the spacetimes described above, we
start from the Lagrangian for the metric (\ref{eq:metric}), 
\begin{equation}
\mathcal{L}=g_{\mu\nu}u^\mu u^\nu + \kappa\,c^2=- a c^2{(u^{0})}^2 + 
a^{-1}{(u^{r})}^2+R^2[{(u^{\theta})}^2+\sin^2\theta {(u^{\varphi})}^2 ] + 
\kappa\,c^2,
\end{equation}
where $u^\mu = \dot{x}^\mu$ is the four velocity and the parameter $\kappa$ assumes the values $1$ or $0$, depending on whether the particle is massive or massless. The line element is spherically symmetric and static, so that the  energy, $E=-\frac{\partial \mathcal{L}}{\partial u^0}$, the azimuthal momentum, $L_\varphi=\frac{\partial \mathcal{L}}{\partial u^\varphi}$, and the total 
angular momentum, $L^2 = \left(\frac{\partial \mathcal{L}}{\partial u^\theta}\right)^2 + \frac{{L_\varphi}^2}{\sin^2\theta}$, are conserved quantities. Explicitly, they have the form:
\begin{eqnarray}
E&=&-\frac{\partial\mathcal{L}}{\partial u^0} = a c^2u^0,\\
L_{\varphi}&=&\frac{\partial\mathcal{L}}{\partial u^\varphi} = R^2 \sin^2\theta u^{\varphi}.
\end{eqnarray}
Since we are only interested in the motion of a single particle (as opposed to a swarm of particles) we can choose the angles such that the orbital plane coincides with the equatorial plane $\theta = \pi/2$, in which case $L_\varphi=L$. In this way, we can express the components of the four-velocity in terms of the conserved quantities $L$ and $E$, and the normalization condition $g_{\mu\nu}\,u^\mu\,u^\nu = -\kappa\,c^2$ yields the radial equation of motion:
\begin{equation}
({u^r})^2+V_{\text{eff}}=\frac{E^2}{c^2},
\end{equation}
with the effective potential
\begin{eqnarray}
V_{\text{eff}}&=&\frac{ a\; L^2}{R^2}+ a\, \kappa\, c^2.
\end{eqnarray}
\begin{figure}[!ht]
		\includegraphics[scale=0.45]{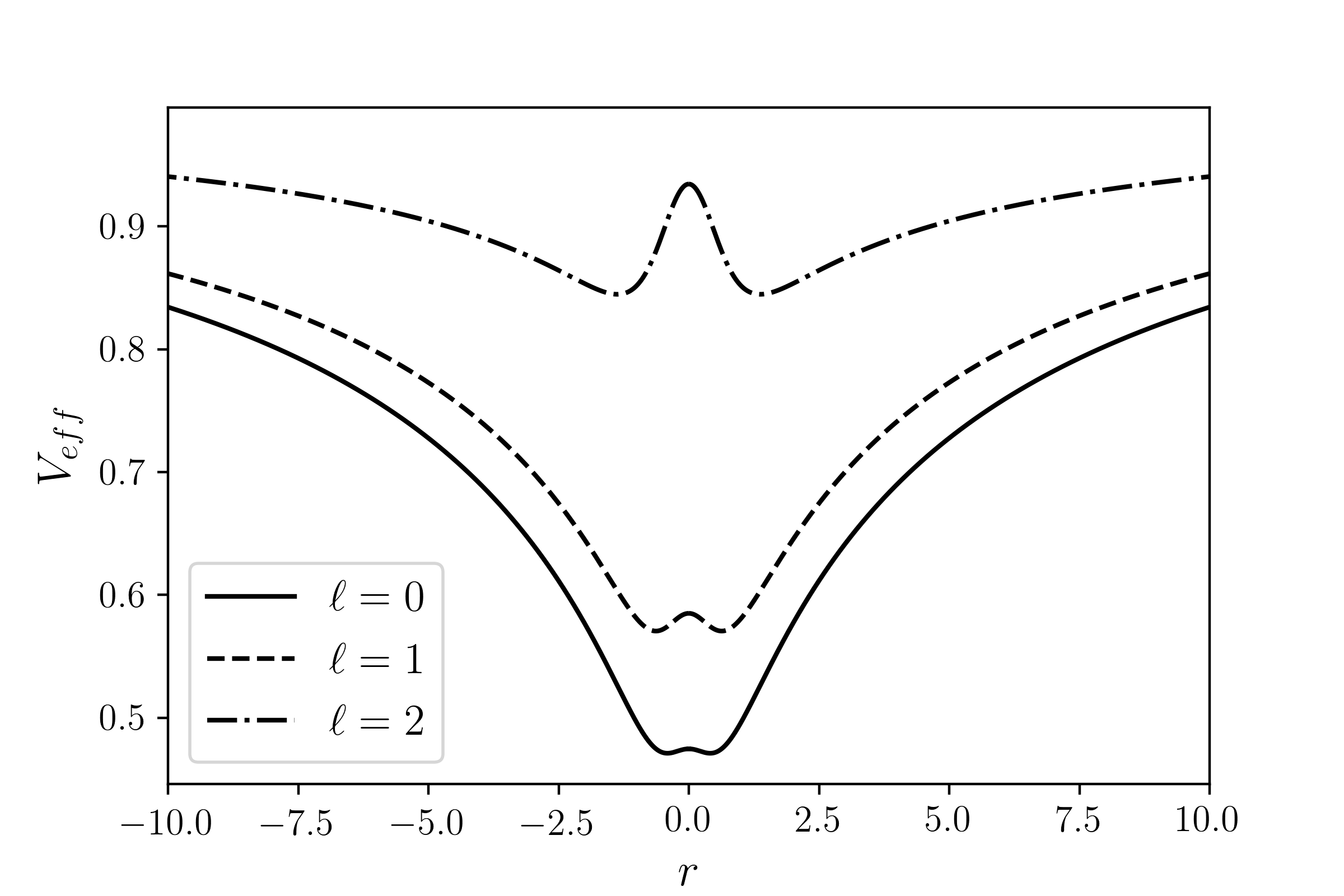} \hspace{1cm}\includegraphics[scale=0.45]{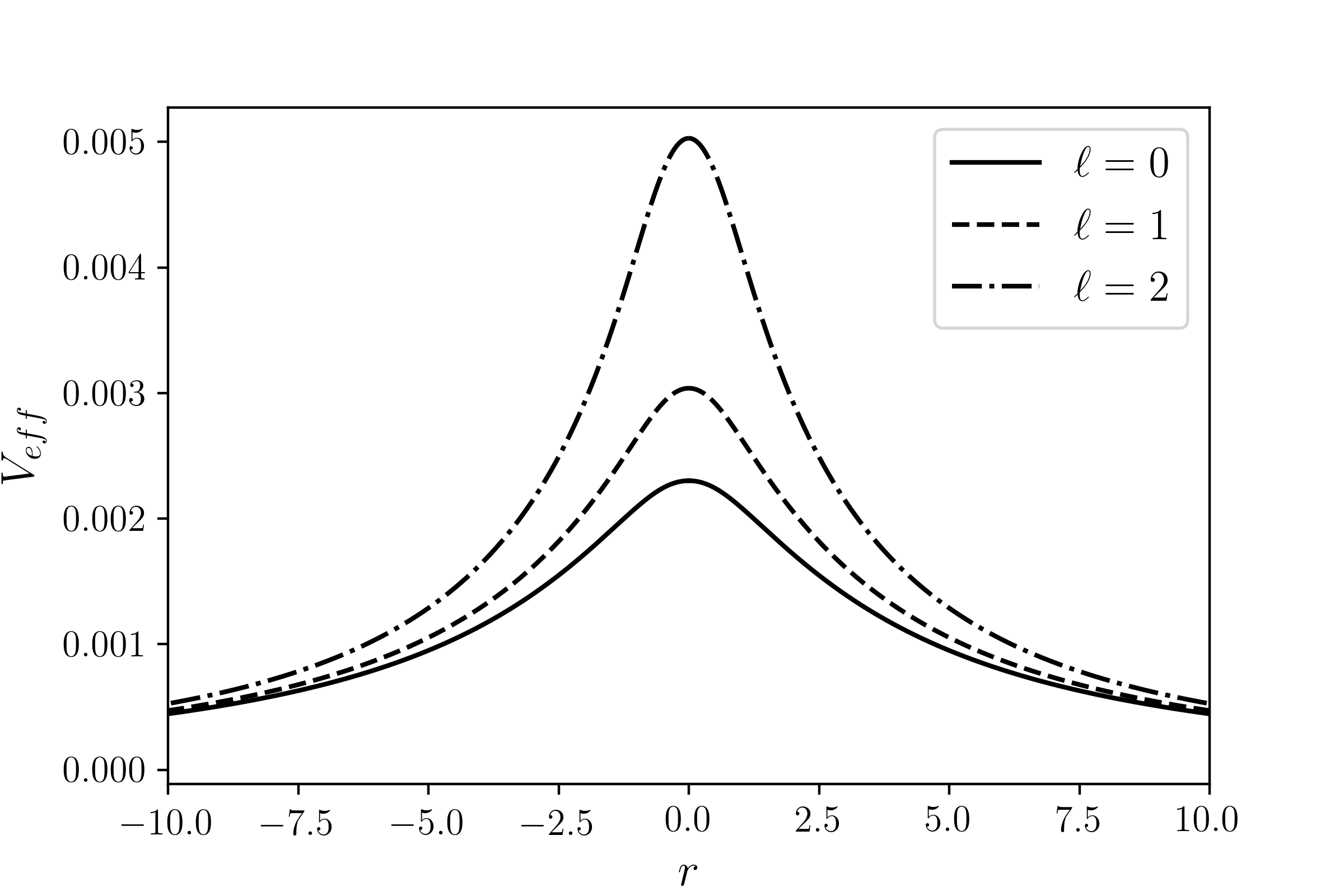}
		\caption{Effective potential for time-like (left panel) and null geodesics (right panel) for $L=\sqrt{\frac{1}{10}}$.}
			\label{fig:potential}
\end{figure}

In Fig.~\ref{fig:potential} we plot $V_{\rm eff}$ for the parameter choices $L=\sqrt{\frac{1}{10}}$, $\omega=0$ and $\Lambda=1.5$ for time-like and a null geodesics. As expected, the term involving $L$ generates an angular momentum barrier, corresponding to a local maximum of the effective potential located at the throat $r = 0$. (Recall from Section~\ref{sec:wormhole} that $a(r)$ has a local maximum while $R^2(r)$ has a local minimum at $r = 0$.) This maximum corresponds to an unstable equilibrium point giving rise to circular unstable particle orbits. For the $\ell=0$ case, and for this value of $L$ and with $\kappa=1$, the effective potential also has a minimum at $r\approx\pm1.34$, which means that bound orbits also exist for this value of $L$.

\begin{figure}[ht]
\includegraphics[width=0.45\columnwidth]{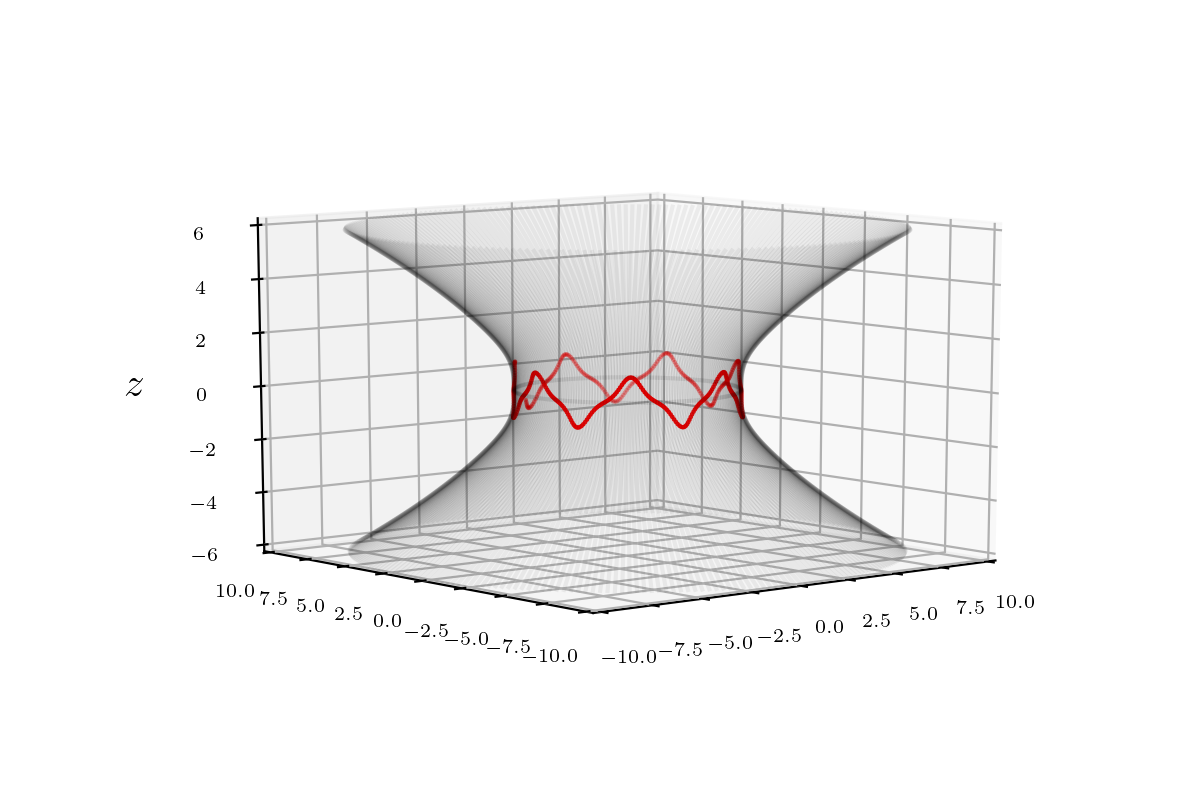}\includegraphics[width=0.45\columnwidth]{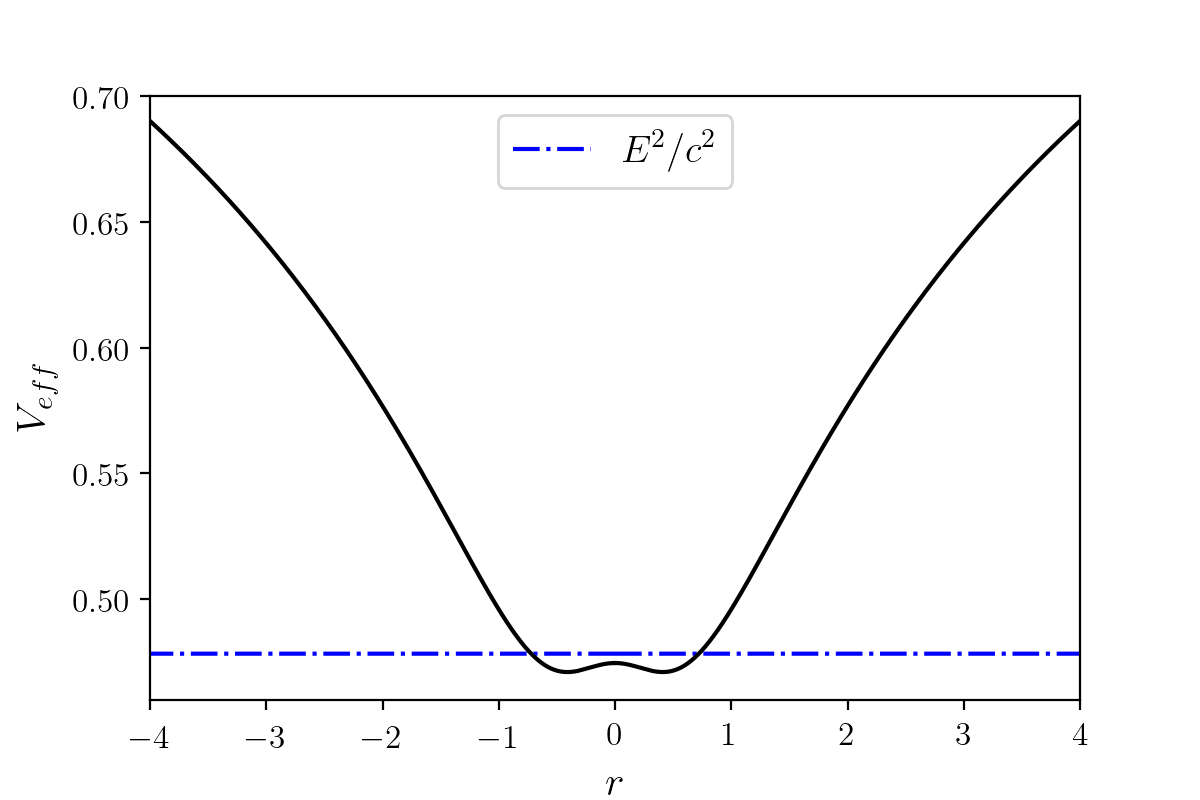}
\\
\includegraphics[width=0.45\columnwidth]{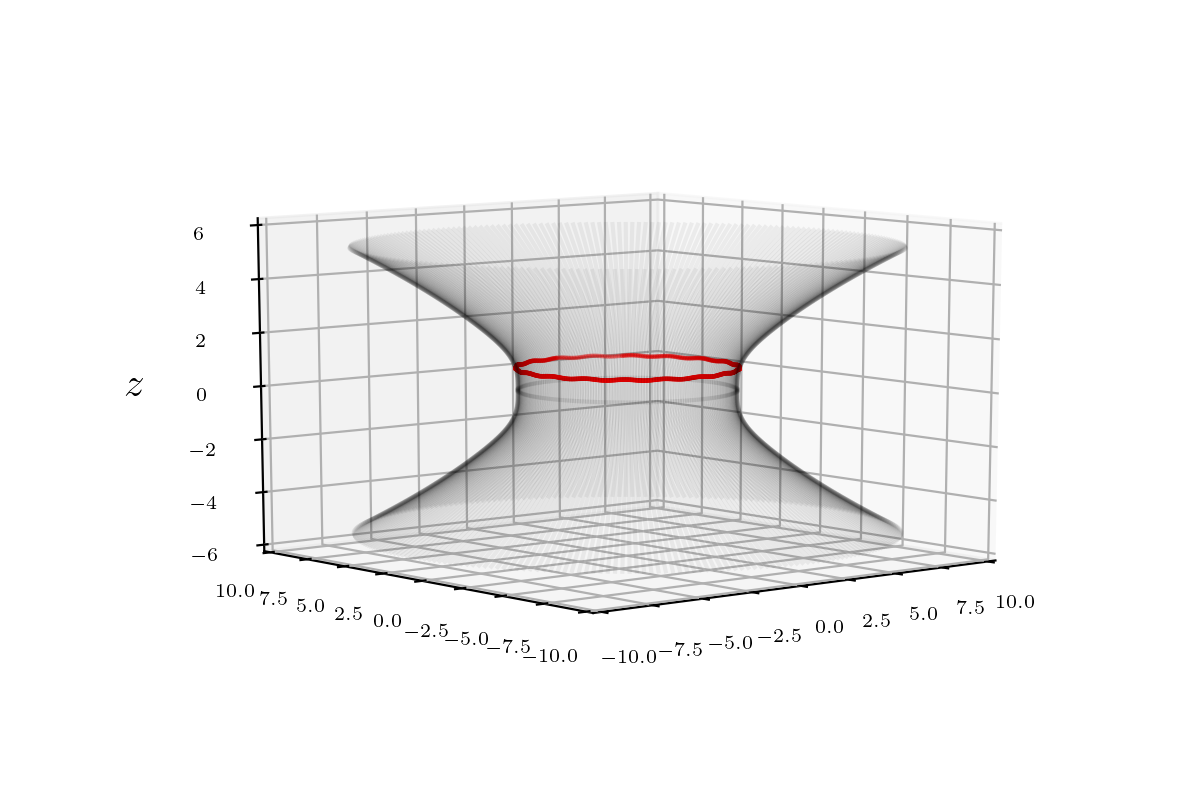}\includegraphics[width=0.45\columnwidth]{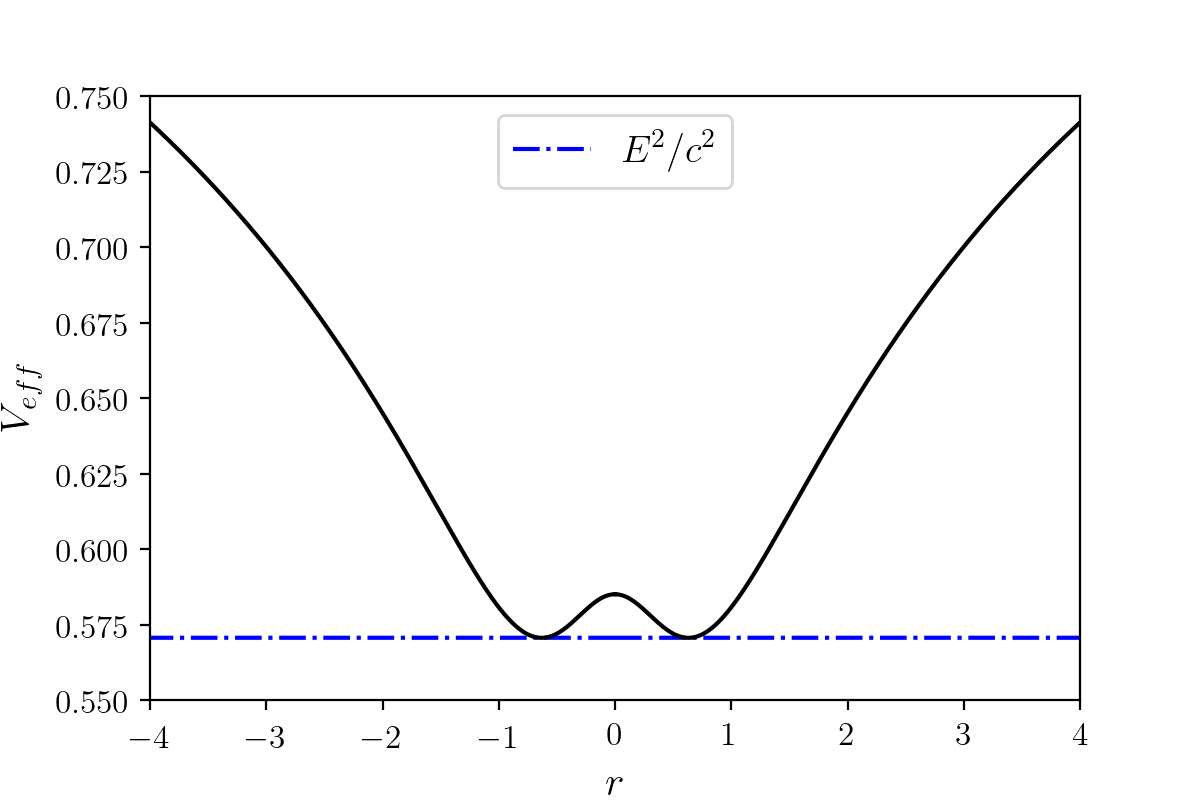}
\\
\includegraphics[width=0.45\columnwidth]{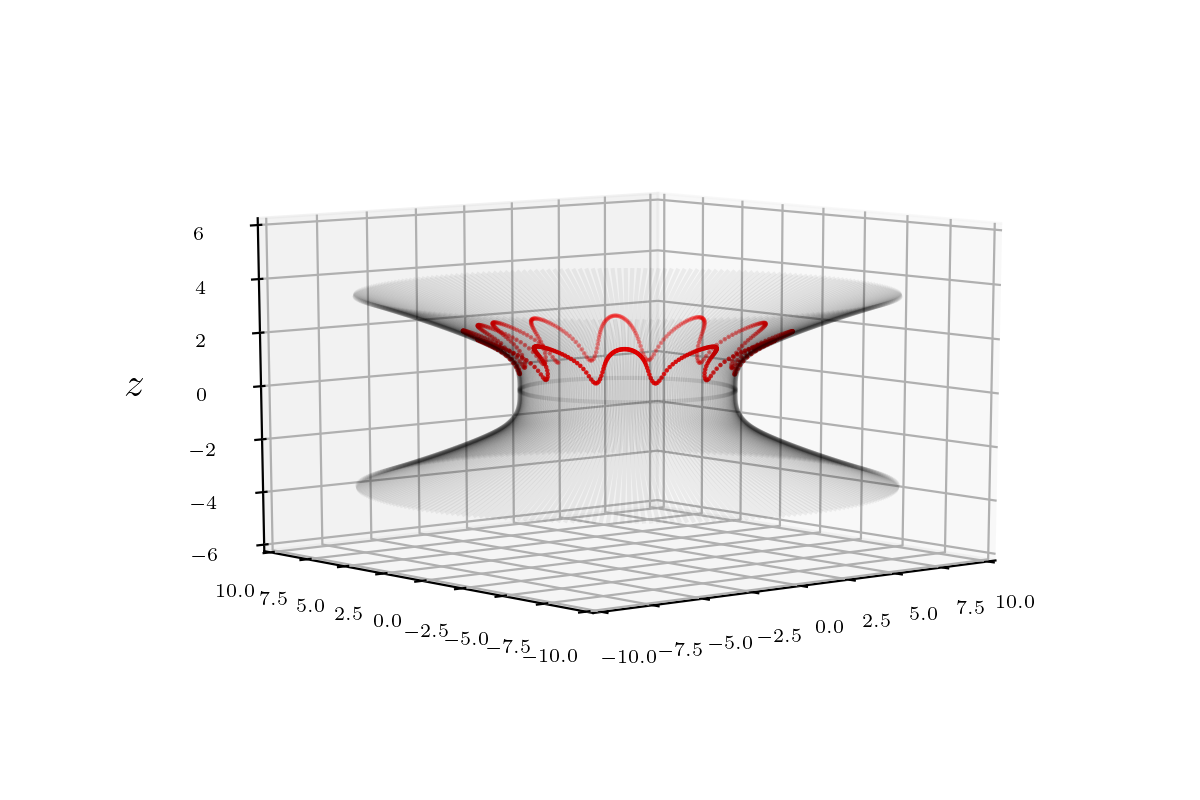}\includegraphics[width=0.45\columnwidth]{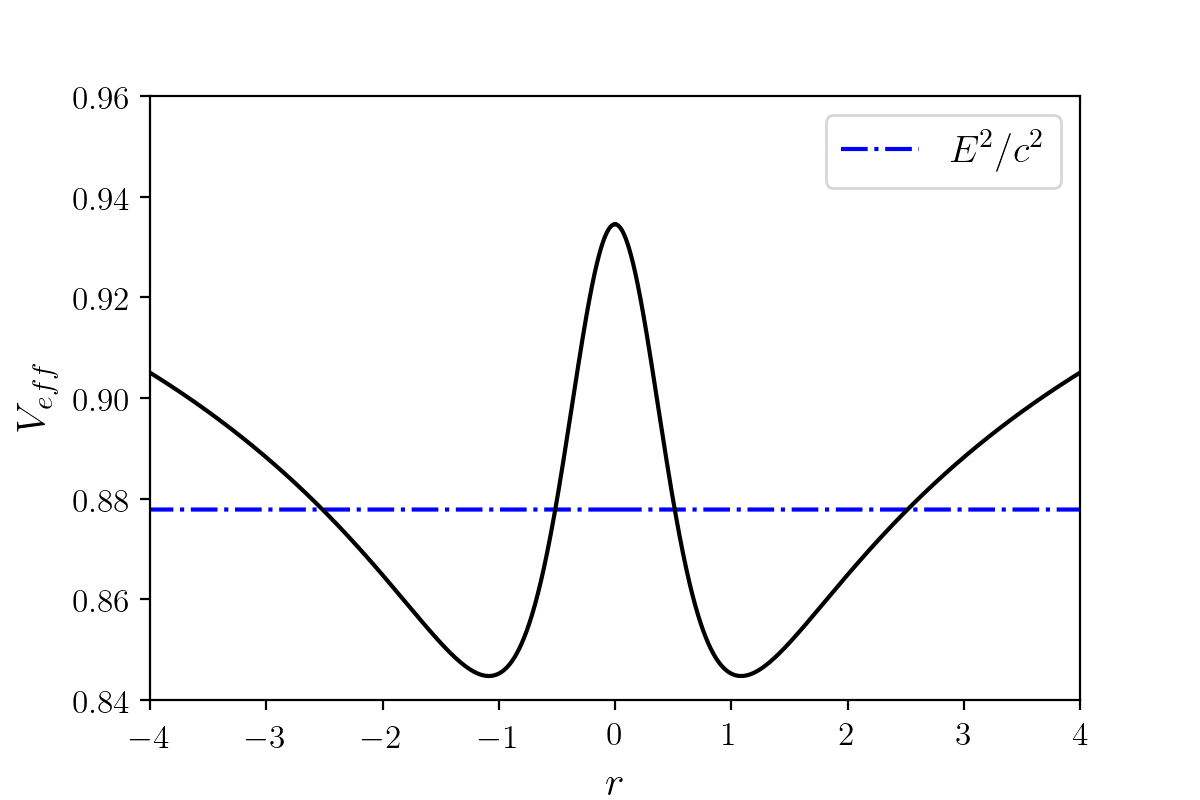}
\caption{Different $\ell=0$ (top), $\ell=1$ (middle) and $\ell=2$ (bottom) 
geodesics for $\kappa=1$, $\omega=0$, $\Lambda=1.5$, $L=\sqrt{1/10}$. On the 
left we plot the motion of the particle in the embedding surface of the wormhole 
and on the right the value of $E$ and $V_{\text{eff}}$.}
\label{fig:geodesics_0}
\end{figure}

In Fig.~\ref{fig:geodesics_0} we plot different geodesics for massive particle with $L=\sqrt{1/10}$ and $\Lambda=1.5$ in the $\ell=0,1,2$ wormholes. Here we picked the same initial conditions in terms of the initial radial velocity $u^r(0)=0$ and initial position $r(0)=0.65$, $\varphi(0)=\pi$ ending up with particles with different energies and qualitatively different motion. As stated above, one can assume without loss of generality that the motion is confined to the equatorial plane $\theta=\frac{\pi}{2}$, so that it can be plotted in the embedding surface. As can be noticed from the plots, the motion is quite interesting and can be understood based on the behavior of the effective potential and the energy level of the test particle.

\begin{center}
	\begin{figure}[!ht]
		\begin{subfigure}{0.9\textwidth}
			\includegraphics[width=0.5\columnwidth]{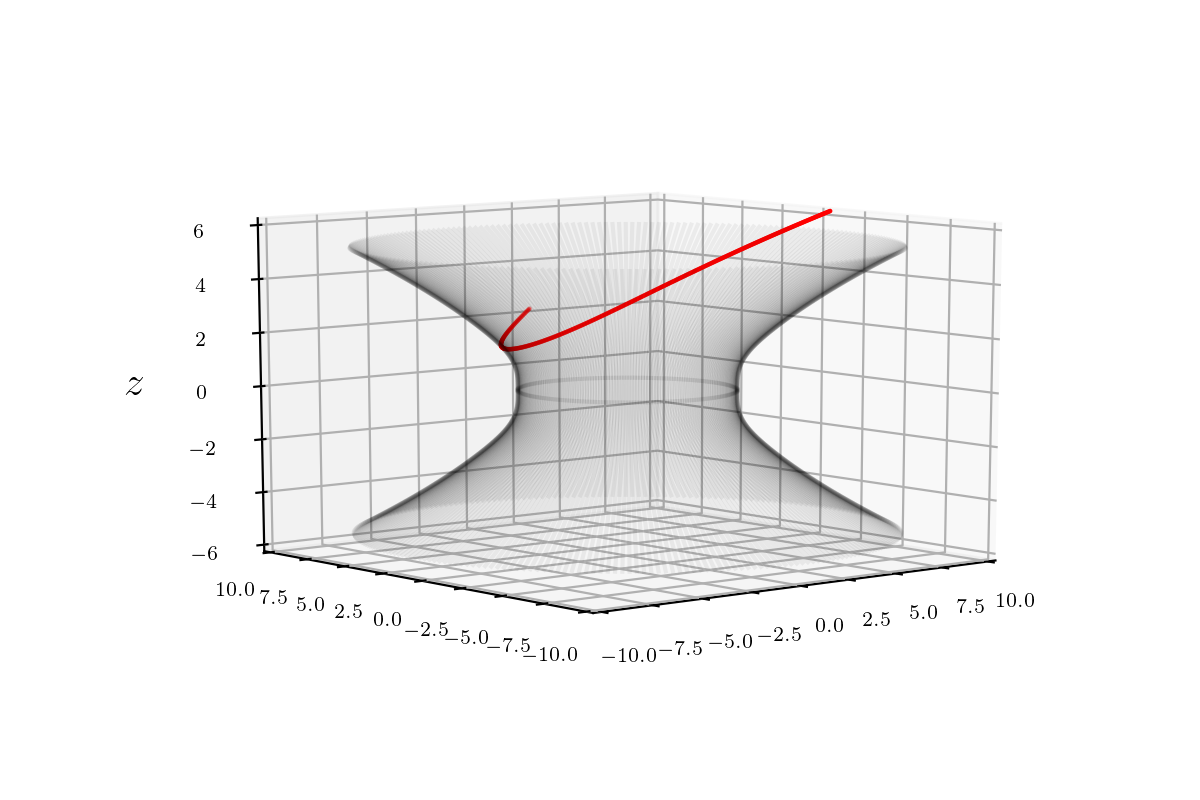}\includegraphics[width=0.45\columnwidth]{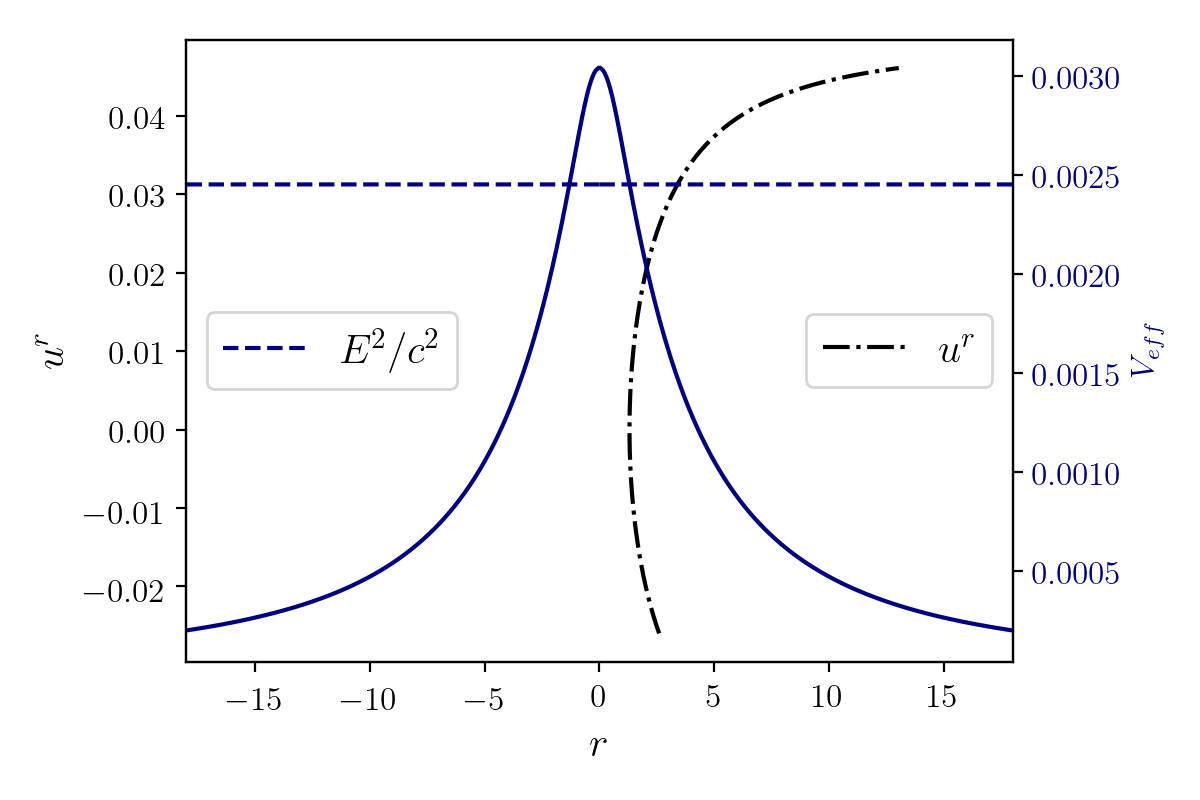}
			\caption{A particle having insufficient energy to pass through the wormhole. The motion starts at $r=2.6$, the velocity decreases until the particle arrives near the throat after which it returns, moving away with increasing speed. In this case the particle stays on the same side of the wormhole and does not cross the throat.} 
			\label{fig:geodesics_vel_a}
		\end{subfigure}
		\\
		\begin{subfigure}{0.9\textwidth}
			\includegraphics[width=0.5\columnwidth]{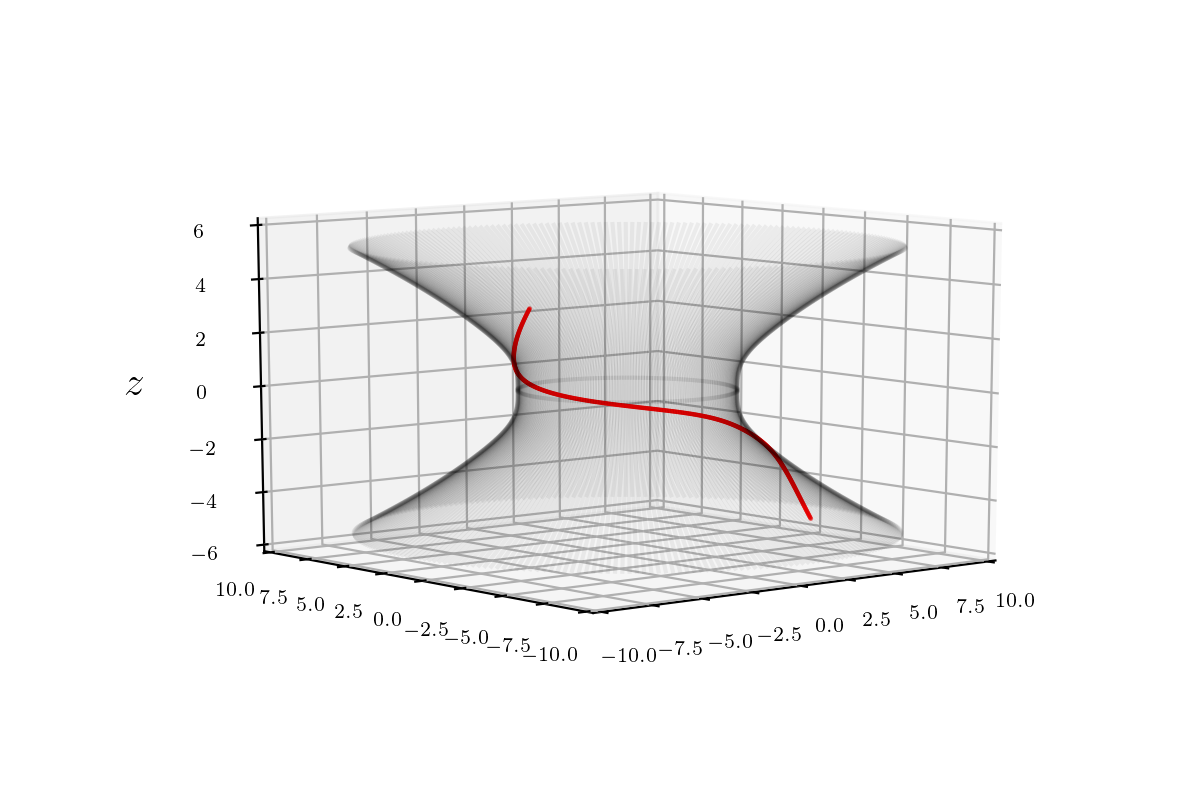}\includegraphics[width=0.45\columnwidth]{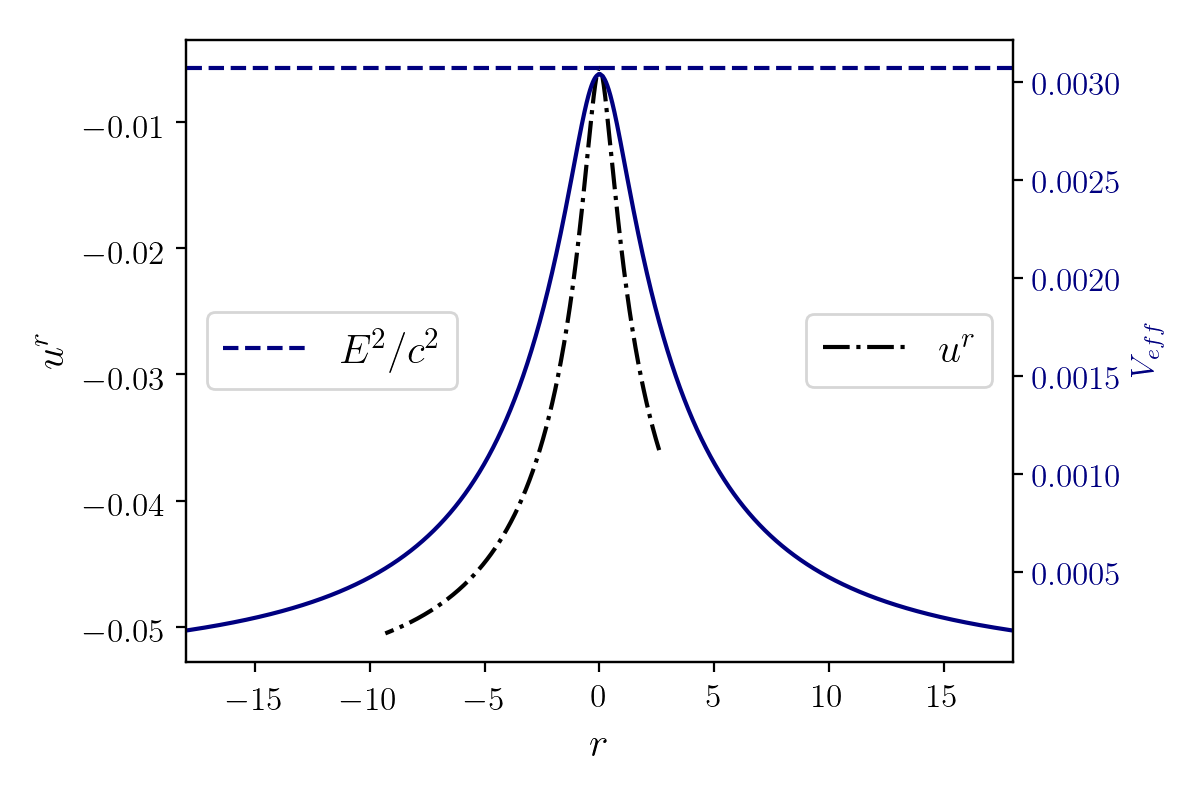}
			\caption{A particle starts its motion at $r=2.6$, traverses the throat, and 
			continues its motion with a growing absolute value of the velocity.
			} \label{fig:geodesics_vel_b}
		\end{subfigure}	
		\caption{Two different $\ell=1$ geodesics for $\kappa=0$, $L=\sqrt{1/10}$, $\omega=0$, $\Lambda=1.5$. We plot the motion with their corresponding $E$ (right y-axis) and $u^r$ (left y-axis) values.} \label{fig:geodesics_vel}
	\end{figure}
\end{center}

In order to further clarify the behavior of the geodesics, in Fig.~\ref{fig:geodesics_vel} we plot the trajectory in the embedding diagram and the radial velocity $u^r$  for two different null geodesics with angular momentum$L=\sqrt{1/10}$ propagating in the $\ell=1$ wormhole. In the first case, shown in Fig.~\ref{fig:geodesics_vel_a}, the particle does not have sufficient energy to traverse the throat so it starts approaching the throat with a decrement of the velocity, reaches a zero radial velocity and resumes its motion going away from the throat. On the other hand, the second example in Fig.~\ref{fig:geodesics_vel_b} shows that, for a particle that has enough energy to pass through the throat, the absolute value of its velocity decreases as it moves towards the throat (from right to left) until it traverses the throat, after which the absolute value of the velocity increases again as the particle moves away from the throat on the other side of the wormhole.

\section{Discussion and concluding remarks}
\label{sec:Conclusions}

We have described how to construct new families of traversable wormhole solutions which are parametrized by a parameter $\ell$, related to the angular momentum of the ghost fields supporting the throat, and discussed its effects on the shape of the geometric functions characterizing the solution, and on the geodesic motion of the corresponding spacetime. These families generalize previous wormhole spacetimes discussed in the literature~\cite{Dzhunushaliev:2017syc, Dzhunushaliev:2008bq} which are recovered from our models by setting $\ell = 0$.

Indeed, we have obtained {\it bona fide} solutions to the Einstein-Klein-Gordon system
and performed a detailed analysis of such solutions, which allowed us to
gain a better understanding on the effects of the new parameter $\ell$. We have been able to establish that its role on the geometric function $a$ (determining the redshift factor), on the curvature scalars and on the density is quite similar to the role played by the frequency $\omega$ characterizing the time-dependency of the field, while its effect on these quantities is opposite to the one generated by the parameter of self-interaction $\Lambda$. Moreover, as can be clearly seen in the plot of the effective potential for time-like geodesics shown in Fig.~\ref{fig:potential}, as the value of $\ell$ grows, the positions of the local minima move farther away from the throat, which is similar to the effect of increasing the angular momentum $L$ of the test particles. In this sense, from the point of view of the test particle, the parameter $\ell$ plays a similar role than its conserved total angular momentum $L$.

It is interesting to point out that the energy density of some of our solutions -- despite of the fact that the stress-energy-momentum violates the null energy condition -- is actually positive close to the throat (but changes its sign as one moves away from it and then converges to zero which is consistent with our asymptotic flatness asymptions). In fact, the construction of a wormhole does not necessarily require measurements of a negative energy density made by static observers, as already indicated in~\cite{Lobo:2004rp}. However, the fact that the null energy condition is violated at the throat implies that such observers also measure a ``superluminal" energy flux. In general, the wormhole solutions discussed in this article possess a much richer structure than the simple, reflection-symmetric Bronnikov-Ellis wormholes, whose energy density is everywhere negative. In particular, the spacetimes discussed here exhibit a rich profile of bumps and wells in their curvature scalars whose precise shape depends on the values of $\ell$ as much as it does on the other parameters.

Indeed, we presented a detailed analysis of the role played by the several parameters in our wormhole solutions, namely the self-interaction term $\Lambda$, the oscillating frequency, $\omega$, and the angular momentum parameter, $\ell$ of the scalar fields. Moreover, we have proved that there are no solutions for which the metric and scalar fields are reflection-symmetric about the throat if $\Lambda = 0$ (see Theorem~1). In this sense, the self-interaction term needs to be included in the action in order to extend the solution space. Actually, we have seen that it plays a smoothing role in the geometric reaction to the ghost matter. Also, we have seen that the effects on the geometry of the self-interaction parameter is opposite to the effects due to the frequency $\omega$. As mentioned previously, the role of the $\ell$ parameter on the geometry is similar to the one generated by the frequency. This fact can be used to obtain real scalar field wormholes, with a new degree of freedom analogous to the case in which the solution space is extended by permitting the scalar field to be complex and harmonic in time. As shown in Theorem~2, all our wormhole solutions are characterized by a single throat whose areal radius is fixed by the parameters and the value of the scalar field at the throat, see Eq.~(\ref{eq:throat}).

We also provided a study of the effects of the parameters on the embedding diagrams visualizing the spatial geometry of the solutions, including the shape of the throat for several relevant cases. Finally, we presented a detailed analysis of the effective potential describing the motion of free-falling test particles as a function of the parameters, and we showed how the potential may present a local maximum at the throat which is surrounded by regions with local minima. Accordingly, we obtained several interesting types of trajectories. Depending on the values of the parameters and on those of the constants of motion (namely, the energy and angular momentum of the particle), we displayed trajectories approaching the throat until they reach a turning point and go back, other trajectories which describe bound motion on either side of the throat, and then we even obtained orbits that are bound but cross the throat repeatedly and keep passing from one side of the Universe to the other; a nice property for a space station!

In the plots of Fig.~\ref{fig:geodesics_vel} we have shown the behavior of the geodesics passing through the throat, we presented the absolute value of the particle's radial velocity and showed that it decreases as the particle approaches the throat until it crosses it after which it increases again as the particle gets further away from the throat. Such behavior is consistent with the interpretation that the reaction of the geometry to the ghost matter is to create bumps in the effective potential, instead of the wells generated by the usual matter. As mentioned at the beginning of Section~\ref{sec:base}, there is no need to invoke negative masses to explain such behavior; it is simpler to imagine that the reaction of the geometry to the ghost matter is to create bumps that the particle have to surmount, consistent with the fact that the absolute value of the velocity decreases as it approaches the throat, and then, goes down the hill.

The new configurations we have found and discussed in this article considerably extend the parameter space describing wormhole solutions of the Einstein-scalar field equations, and they provide a large arena that offers the possibility to further study the intriguing properties of wormhole spacetimes, including the relation between the properties of exotic matter and their geometry. While it has been shown that the solutions with $\ell = 0$ are linearly unstable~\cite{Gonzalez:2008wd,Gonzalez:2008xk,Dzhunushaliev:2017syc}, there is hope that such a large arena may contain a set of parameter values with $\ell > 0$ describing stable wormholes or unstable wormholes with a very large timescale associated to their instability, a question that will be discussed in a future work.
\vspace{-5mm}


\acknowledgments
\vspace{-5mm}
We thank Fabrizio Canfora for pointing out to us the existence of traversable wormhole geometries in the absence of exotic matter.
This work was partially supported by 
DGAPA-UNAM through grants IN110218 and IA101318 and by the CONACyT Network Project No. 294625 
``Agujeros Negros y Ondas Gravitatorias". This work has further been supported 
by the European Union's Horizon 2020 research and innovation (RISE)
program H2020-MSCA-RISE-2017 Grant
No. FunFiCO-777740.
BC and VJ acknowledge support from CONACyT. OS was partially supported by a CIC grant to Universidad Michoacana.

\newpage

\bibliographystyle{unsrt}
\bibliography{biblio}


\end{document}